\documentclass[referee,sn-basic]{sn-jnl}

\usepackage{graphicx}%
\usepackage{tabularx}
\usepackage{multirow}%
\usepackage{amsmath,amssymb,amsfonts}%
\usepackage{amsthm}%
\usepackage{mathrsfs}%
\usepackage[title]{appendix}%
\usepackage{xcolor}%
\usepackage{textcomp}%
\usepackage{manyfoot}%
\usepackage{booktabs}%
\usepackage{algorithm}%
\usepackage{algorithmicx}%
\usepackage{algpseudocode}%
\usepackage{listings}%
\usepackage{graphicx}
\usepackage{subcaption} 
\usepackage{caption}
\usepackage{float} 
\usepackage{lmodern}
\usepackage{anyfontsize}
\usepackage{newtxtext,newtxmath}
\begin{document}

\title[clustering]{Spatiotemporal clustering of GHGs emissions in Europe: exploring the role of spatial component}


\author*[1]{\fnm{Caterina} \sur{Morelli}}\email{c.morelli12@campus.unimib.it}

\author[1]{\fnm{Paolo} \sur{Maranzano}}\email{paolo.maranzano@unimib.it}

\author[2]{\fnm{Philipp} \sur{Otto}}\email{philipp.otto@glasgow.ac.uk}

\affil*[1]{\orgdiv{Department of Economics, Management and Statistics (DEMS)}, \orgname{University of Milano-Bicocca}, \orgaddress{\street{Piazza dell'Ateneo Nuovo n.1}, \city{Milano}, \postcode{20126}, \country{Italy}}}


\affil[2]{\orgdiv{School of Mathematics and Statisitcs}, \orgname{University of Glasgow}, \orgaddress{\street{University Place}, \city{Glasgow}, \postcode{G12 8QQ}, \state{Scotland}, \country{United Kingdom}}}



\abstract{
In this study, we propose a novel application of spatiotemporal clustering in the environmental sciences, with a particular focus on regionalised time series of greenhouse gases (GHGs) emissions from a range of economic sectors. Utilising a hierarchical spatiotemporal clustering methodology, we analyse yearly time series of emissions by gases and sectors from 1990 to 2022 for European regions at the NUTS-2 level. While the clustering algorithm inherently incorporates spatial information based on geographical distance, the extent to which space contributes to the definition of groups still requires further exploration. To address this gap in the literature, we propose a novel indicator, namely the Joint Inertia, which quantifies the contribution of spatial distances when integrated with other features. Through a simulation experiment, we explore the relationship between the Joint Inertia and the relevance of geography in exploiting the groups structure under several configurations of spatial and features patterns, providing insights into the behaviour and potential of the proposed indicator. The empirical findings demonstrate the relevance of the spatial component in identifying emission patterns and dynamics, and the results reveal significant heterogeneity across clusters in trends and dynamics by gases and sectors. This reflects the heterogeneous economic and industrial characteristics of European regions. The study highlights the importance of the spatial and temporal dimensions in understanding GHGs emissions, offering baseline insights for future spatiotemporal modelling and supporting more targeted and regionally informed environmental policies.
}


\maketitle

\section{Introduction}\label{Sec_Intro}

The growing amount of spatiotemporal data available poses increasingly complex and intriguing challenges to researchers. In the environmental field, in particular, new technologies enable the acquisition of georeferenced data, in which spatial information is incorporated in different ways, providing a wealth of opportunities for detailed analysis. Remote sensing, GPS devices, mobile sensors, and satellite imagery generate massive spatiotemporal datasets, capturing environmental phenomena at unprecedented temporal and spatial scales. Numerous studies have focused on developing advanced spatiotemporal statistical models to predict and understand pollutant concentration behavior in different geographical areas and over time \citep{Calculli2015,Najafabadi2020,Taghavi-Shahri2020,Maranzano2023,Otto2024245} and other papers examine the effects of specific events \citep{Maranzano2022343,Maranzano2024147} or policies \citep{Maranzano2020} on pollution levels.
However, while the concentration of pollutants in the air has been widely studied, there remains a critical gap in the analysis of greenhouse gas (GHGs) emissions, particularly those directly resulting from human and economic activities. This lack of focus on the sources of emissions may constrain our understanding of the fundamental drivers and spatial distribution of environmental degradation. While air pollution is undoubtedly influenced by GHGs emissions, it is also shaped by various other factors, such as the morphological characteristics of the terrain, including elevation, land use, and proximity to natural barriers like mountains or water bodies. These geographical features can significantly affect the dispersion, accumulation, and concentration of pollutants in the atmosphere, adding complexity to the relationship between human activities, GHGs emissions, and localized air quality. To encourage researchers to look at air pollution from a new and challenging perspective, spatiotemporal cluster analysis offers a powerful approach to start bridging this gap, enabling the identification of spatial and temporal patterns of GHGs emissions. 

Through clustering techniques, large datasets can be grouped based on the similarity of their spatial and temporal attributes, allowing researchers to discover trends, anomalies, and insights that are crucial for addressing complex environmental challenges \cite{Birant2007208}. Spatiotemporal clustering enables the extraction of spatial and temporal relationships or similar patterns from data that is distributed over both time and space. In the context of GHGs emissions, such analysis can provide deeper insights into the spatiotemporal distribution of emissions across regions, highlight emission hotspots, and reveal how these emissions evolve over time in response to various factors such as industrial activity and economic changes.


The application of spatiotemporal clustering has shown promise in various areas such as seismology \citep{Georgoulas20134183,Wang2006263}, traffic management \citep{Jeung20081068,Tietbohl2008863} and risk management \citep{Lee20129623}.
In the realm of environmental science, it has been used for monitoring deforestation and vegetation distribution \citep{Agrawal2016388}, water quality and sea water temperature \citep{Alatrista-Salas2015127,Birant2007208}, tracking wildlife migration and animal behavior \citep{Kalnis2005364,Vieira2009286}, identifying patterns of environmental impact evaluation of firms \citep{Morelli2024} and the interaction between climate change awareness and climate-related characteristics in different countries \citep{Zammarchi2024}.

The statistical literature proposes several methodologies of spatiotemporal cluster analysis, which are reported in a literature review proposed by \cite{Ansari20202381} and classified according to a taxonomy on different data types. Among all the proposed methodologies, most papers exploit density-based spatial clustering algorithm as proposed by \citep{Ester1996226} named Density-Based Spatial Clustering of Applications with Noise (DBSCAN), and its extensions \citep{Tietbohl2008863,Rocha2010114,Georgoulas20134183,Chen20152575,Birant2007208}. Density-based clustering is able to detect clusters of arbitrary shape and different sizes without setting the number of clusters as the input parameters, but they are very sensitive to other input parameters (i.e., radius, minimum number of points, threshold).

In this paper, we rely on spatiotemporal hierarchical clustering as proposed by \citep{Chavent2018} and extended to the spatiotemporal case by \citep{Morelli2024}. Although this approach is relatively recent, it has been widely adopted in the literature \citep{Jaya2019,Bucci2023214,Mattera2023,Deb2023,Morelli2024,Zammarchi2024}, due to its ability to group observed units without imposing assumptions on data distribution or cluster shapes. Its flexibility is another key advantage, as it relies on a hierarchical algorithm, enabling the inclusion of any type of data by simply calculating the dissimilarity between observations using an appropriate distance measure. Another important advantage is that the version proposed by \cite{Morelli2024} allows a very large number of diminutions to be input (in fact, it is possible to include time series of several variables), and the algorithm will assign different weights to the dissimilarity matrices so that the distances between observations can be explained as much as possible. So, the algorithm will only take into account information that is useful for identifying clusters. In this context, we propose a measure designed to evaluate the contribution of one dissimilarity matrix relative to others, with the specific goal of assessing the role of the spatial component when combined with distances calculated from other variables. After providing a theoretical explanation of this measure, supported by examples of hypothetical scenarios, we demonstrate its behavior through a simulation experiment. This experiment explores how the measure responds to variations in the spatial distance between clusters and their degree of overlap in space.

This measure is fundamentally important, not only in our specific application but also in other contexts. Given that cluster analysis is a form of exploratory analysis applied to data whose characteristics are not yet fully understood, it is not appropriate to a priori include the spatial component (or other variables describing distinct phenomena) and force the algorithm to consider all input information equally. Instead, it is crucial to understand the actual role of spatial information, especially when it is associated with a specific phenomenon that has not yet been examined through a spatiotemporal approach. 

The contributions of this paper are twofold, addressing both methodological innovation and practical application in the environmental domain. From a methodological perspective, we introduce a novel measure to assess the role of the spatial component in clustering, particularly when it is combined with dissimilarities derived from other variables. This measure provides a valuable tool for understanding the influence of spatial information on clustering outcomes and serves as a foundation for future analyses involving spatiotemporal data. From an application standpoint, we approach the environmental issue of GHGs emissions from a fresh perspective, offering insights into the patterns and dynamics of emissions across regions and over time. A key objective is to evaluate the relevance of spatial information in identifying these patterns, thereby contributing to a deeper understanding of GHGs emissions and their spatial-temporal distribution.

The relevance of these results extends to both policymakers and statisticians. For policymakers, the insights gained from this analysis are instrumental in formulating localized strategies to mitigate GHGs emissions, enabling more targeted and effective interventions. For statisticians in the environmental field, this study provides a baseline analysis that highlights critical factors to consider when modeling GHGs emissions data. Specifically, it underscores the importance of incorporating spatial and temporal dimensions and offers guidance on whether spatiotemporal models are necessary for accurate representation and prediction. Together, these contributions lay the groundwork for more robust and nuanced approaches to addressing pressing environmental challenges.

The rest of the paper is structured as follows: in Section \ref{sec_method}, we describe the methodology proposed by \cite{Chavent2018} and \cite{Morelli2024}, in Section \ref{sec_jointInertia}, we provide the methodological contribution of the paper, in Section \ref{sec_sim} we present the simulation experiment on the performance of spatiotemporal clustering algorithm and the measure to assess the role of the spatial component and in \ref{sec_app} we discuss the empirical findings, thus the clusters obtained from the algorithm and the contribution of the spatial distances.


\section{Spatiotemporal hierarchical clustering}\label{sec_method}
In this section, we briefly recall the baseline methodology proposed by \cite{Chavent2018}, in which two dissimilarity matrices, one containing spatial information and the other containing clustering features, are combined to cluster the units under some geographical constraint. Then, we describe the extension proposed by \cite{Morelli2024}, which considers a spatiotemporal framework and adds a novel algorithm for the identification of the weights for the spatial and non-spatial matrices. Finally, we propose a metric that allows us to evaluate the role of the spatial component in explaining the dissimilarity of the data and, thus, the relevance of the spatial information with respect to the features used in the cluster analysis.

\subsection{Baseline methodology: spatial hierarchical clustering}
Let $D_0 = [d_{0,ij} ]_{i,j=1,\dots,n}$ and $D_1 = [d_{1,ij} ]_{i,j=1,\dots,n}$ refer respectively to any distances matrix of variables and the spatial distances matrix considering a sample of $n$ units. Also, let $w_i$ with $i = 1,\dots, n$ be the weight of the $i$-th cross-sectional unit to be clustered. When no prior information is available, the weight is commonly set to $w_i=1/n$. In order to account for different measurement scales, it is necessary to scale $D_0$ and $D_1$ with respect to their maximum values so that the clustering distances across observations take values between $0$ and $1$. 
In the spatial clustering proposed by \cite{Chavent2018}, the matrices $D_0$ and $D_1$ are combined through a convex combination with parameter $\alpha$, which provides the weight of geographical information in determining the clusters. When the mixing parameter is null, that is, when $\alpha = 0$, in the linear combination of the dissimilarity matrices $D(\alpha)=(1-\alpha)D_0+\alpha D_1$, the geographical dissimilarities are not taken into account, while when $\alpha = 1$ the features distances are ignored, and the clusters are defined only based on the geographical distances.

Once the combined matrix $D$ is obtained, the Ward hierarchical clustering algorithm \citep{Ward1963236} is applied to cluster the units. In its original formulation, the starts with an initial partition into $n$ clusters of singletons, and at each step, the algorithm aggregates the two clusters such that the new partition has minimum within-cluster inertia. In our context, given a partition $\mathcal{P}_K^{\alpha}=(\mathcal{C}_1^{\alpha}, \dots, \mathcal{C}_K^{\alpha})$ into $K$ clusters, the mixed pseudo-inertia for cluster $\mathcal{C}_K^{\alpha}$ is defined as the convex combination of the sum of square dissimilarities in $D_0$ and $D_1$ across the units belonging to cluster $\mathcal{C}_K^{\alpha}$. The corresponding mixed pseudo-inertia, denoted as $I(\mathcal{C}_K^{\alpha})$, is computed as
\begin{equation}
I(\mathcal{C}_K^{\alpha})=(1-\alpha)\sum_{i \in \mathcal{C}_K} \sum_{j \in \mathcal{C}_K} \frac{w_i w_j}{2 \sum_{i \in \mathcal{C}_K} w_i} {d_{0,ij}}^2 + \alpha \sum_{i \in \mathcal{C}_K} \sum_{j \in \mathcal{C}_K} \frac{w_i w_j}{2 \sum_{i \in \mathcal{C}_K} w_i} {d_{1,ij}}^2    
\end{equation}

The overall pseudo-within-cluster inertia of the partition to be minimized is obtained as the sum of the pseudo-inertia of each cluster and, thus, is computed as follows
\begin{equation}
W_\alpha(\mathcal{P}_K^{\alpha})=\sum_{k=1}^{K} I_\alpha(\mathcal{C}_k^{\alpha}).    
\end{equation}
The algorithm aims to identify the cluster partition such that units in the same clusters have a lower total square dissimilarity. Before proceeding, we point out that pseudo-inertia is a generalization of inertia, where the dissimilarities can be either Euclidean or non-Euclidean. From here on, with a slight abuse of notation, we will refer to inertia in substitution of the pseudo-inertia.

The main issue in the hierarchical clustering methodology concerns the choice of the mixing parameters $\alpha$ and the number of clusters $K$. \cite{Chavent2018} suggest setting a prior value for $K$ and then finding $\alpha$ such that it allows to explain the same proportion of the dissimilarities from both matrices, with respect to the cases in which the clusters are obtained considering only the feature matrix or the spatial matrix. In order to clarify their approach, we recall the notion of the proportion of the total inertia explained by partition $\mathcal{P}_K^{\alpha}$ in $D_0$ and $D_1$ for $K$ clusters, which is computed as
\begin{equation}
Q_{D_0}(\mathcal{P}_K^{\alpha})=1-\frac{W_{D_0}(\mathcal{P}_K^{\alpha})}{W_{D_0}(\mathcal{P}_1)} \qquad Q_{D_1}(\mathcal{P}_K^{\alpha})=1-\frac{W_{D_1}(\mathcal{P}_K^{\alpha})}{W_{D_1}(\mathcal{P}_1)} \,   
\end{equation}
Specifically, $Q_{D_{0}}(\mathcal{P}_K^{\alpha})$ quantifies the proportion of inertia from features dissimilarities (i.e., $W_{D_0}(\mathcal{P}_1)$) explained by partition $\mathcal{P}_K^{\alpha}$, while $Q_{D_{1}}(\mathcal{P}_K^{\alpha})$ quantifies the amount of geographical inertia (i.e., $W_{D_1}(\mathcal{P}_1)$) explained by partition $\mathcal{P}_K^{\alpha}$.

To account for potential scale issues in $Q_{D_{0}}(\mathcal{P}_K^{\alpha})$ and $Q_{D_{1}}(\mathcal{P}_K^{\alpha})$, the $Q_{\beta}(\mathcal{P}_K^{\alpha})$ metrics are then normalized with respect to the baseline case of purely-geographical or purely-features clustering, that is, by computing the following ratios:
\begin{equation}
\tilde{Q}_{D_0}(\mathcal{P}_K^{\alpha})=\frac{Q_{D_0}(\mathcal{P}_K^{\alpha})}{Q_{D_0}(\mathcal{P}_K^0)} \qquad \tilde{Q}_{D_1}(\mathcal{P}_K^{\alpha})=\frac{Q_{D_1}(\mathcal{P}_K^{\alpha})}{Q_{D_1}(\mathcal{P}_K^1)}.
\label{eq:norm_prop_inertia}
\end{equation}
This relative formulation allows for a straightforward interpretation of the values. For a given $K$ and a given mixing parameter $\alpha$, the formulas are expressing the percentage of the explained proportion of inertia in features (or spatial) dissimilarity matrix obtained by using a mixture of features and geographical information to generate the partition $\mathcal{P}_K^{\alpha}$ (i.e., $Q_{D_0}(\mathcal{P}_K^{\alpha})$ (or $Q_{D_1}(\mathcal{P}_K^{\alpha})$)) with respect to the proportion of inertia that would be explained by only using features (or spatial) dissimilarities to generate the partition $\mathcal{P}_K^{0}$ (or $\mathcal{P}_K^{1}$) in $K$ clusters (i.e., $Q_{D_0}(\mathcal{P}_K^0)$ (or $Q_{D_1}(\mathcal{P}_K^1)$)). 
For a deeper understanding and comprehensive discussion about the properties of these quantities, we refer the readers to Section 3 in \cite{Chavent2018}.

\cite{Chavent2018} suggest to choose $\alpha$ such that the normalized proportion of the explained pseudo inertia from $D_0$ and $D_1$ are as similar as possible, that is,
\begin{equation}
    min_{\alpha} |\tilde{Q}_{D_0}(\mathcal{P}_K^{\alpha}) - \tilde{Q}_{D_1}(\mathcal{P}_K^{\alpha})| .
\end{equation}
In other words, this means finding $\alpha$ such that the proportion of explained pseudo inertia attributed to the features and the spatial information is as close as possible. Specifically, this ensures that the proportion of inertia explained when using only features or only spatial information to generate the partition is similar. To select the number of clusters $K$, \cite{Chavent2018} suggest to choose it a priori. Similarly, \cite{Mattera2023} set an initial number of clusters $K_0$ considering the partition associated with $D_0$, then they determine $\alpha$ as in \cite{Chavent2018}, and finally they define the optimal number of clusters based on the combined dissimilarity matrix. 


Notice that this selection method does not always allow to identify $\alpha$ such that it captures the highest possible overall dissimilarity in the data. To address such drawback, \cite{Jaya2019} start finding $\alpha$ according to \cite{Mattera2023}, and they end up choosing a different value for the mixing parameter in order to explain better the proportion of inertia in one matrix, with a relatively small reduction of the proportion of inertia from the other matrix.

Moreover, the algorithm is induced to select a combination of matrices such that both $D_0$ and $D_1$ are always included, without knowing if and how much the geographic component is really relevant in the cluster formation. In other words, we do not know what role spatial distance plays in the representation of the phenomenon we are examining. This is a very important point since cluster analysis is used as exploratory analysis, so it could be applied when we start studying new or little-analyzed variables in literature, and we often do not have a large amount of previous knowledge. It would, therefore, be preferable to apply a methodology that allows the geographical distance between observations to be considered only when providing some actual improvements in the clustering results.

\subsection{Spatiotemporal hierarchical clustering}
In \cite{Morelli2024}, the authors introduced an enhanced version of spatial hierarchical clustering that allows the use of more than two dissimilarity matrices. Also, they identify the optimal combination of these matrices which maximizes the gain induced by using a mixed approach instead of a purely spatial and non-spatial approach in terms of proportion of explained inertia within the clusters without forcing the algorithm to include all the input matrices in the final combination. Such principle holds for both the spatial and non-spatial information.

The need to incorporate multiple dissimilarity matrices arose from our objective of analyzing a dataset consisting of time series for more than one variable and the geographic coordinates of individual units. Indeed, the approach proposed by \cite{Chavent2018} allows for the construction of a single matrix containing the Euclidean distances of variables observed at a given moment or a matrix representing the distances between time series for a single variable \citep{Bucci2023214, Deb2023, Mattera2023}. When dealing with multiple time series, it becomes necessary to compute a dissimilarity matrix for each variable observed over a given period. Therefore, the challenge lies in combining more than two matrices. Notice that the methodology is not limited to spatiotemporal data but can be applied to any scenario where high-dimensional data requires the combination of more than two dissimilarity matrices. Nevertheless, this approach has been developed specifically to address the need for identifying patterns and dynamics in phenomena where both temporal evolution and geographic location are critical.

Let $D_p = [d_{p,ij}]_{p=1,\dots, P-1; i,j=1,\dots,n}$ being the dissimilarity matrices containing the distances across time series for $P$ variables for $n$ units. Let us assume that the first $P-1$ matrices are associated with non-spatial features or time series and let $D_P = [d_{P,ij}]_{i,j=1,\dots,n}$ refer to the spatial (or geographical) dissimilarity matrix across the $n$ units, which is typically computed as geodetic distances between coordinates. Moreover, we recall that dissimilarity matrices are normalized with respect to their maximum value. The main issue here is to find a suitable combination of the two clustering hyperparameters, namely the number of group $K$, and the weighting vector $\underline{\alpha}=[\alpha_p]_{p=1,2,\dots,P} = (\alpha_1,\alpha_2,\ldots,\alpha_p,\ldots,\alpha_P)$ with $\alpha_p>0 \quad \forall p=1,\ldots,P$ and such that $\sum_{p=1}^P\alpha_p=1$. Notice that, although the grid can be either regular or irregular, we suggest using a regular grid with constant step $\Delta \alpha$. Let us define a range of number of clusters from $K=1$ to $K=K_{max}$ and for each $K=1,\ldots,K_{max}$ consider all the possible combinations for a grid of $\underline{\alpha}$. For each combination of $K$ and $\underline{\alpha}$ we compute the corresponding convex linear combination of dissimilarity matrices $D(\underline{\alpha})=\sum_{p=1}^P \alpha_pD_p$ and we identify the clustering partition $\mathcal{P}_K^{\underline{\alpha}}$ according to the Ward hierarchical algorithm.

Following the criterion defined in \cite{Morelli2024}, for every $K=1,2,\ldots,K_{max}$ we choose the optimal weighting vector $\underline{\alpha} = \underline{\alpha}^*$ which maximizes the weighted average of the explained mixed inertia\footnote{We refer the reader to the detailed discussion in \cite{Morelli2024} about the analytics and the interpretation of the weighted average of explained mixed pseudo inertia, in particular to its link with the gain in using a mixed approach compared to the purely non-spatial or purely spatial cases.} induced by the partition, given by
\begin{equation}
\bar{Q}(\mathcal{P}_K^{\underline{\alpha}})=1-\frac{\sum_{p=1}^P W_{D_p}(\mathcal{P}_K^{\underline{\alpha}})}{\sum_{p=1}^P W_{D_p}(\mathcal{P}_1^{\underline{\alpha}})}.    
\end{equation}
Its value ranges between $0$ and $1$, reaching its maximum when partition $\mathcal{P}_K^{\underline{\alpha}}$ allows us to explain the total dissimilarity in each matrix. The optimization step conditioning on $K$ is iterated across a user-defined range of number of clusters $K=1,2,\dots,K_{max}$ evaluated at the corresponding optimal $\underline{\alpha}_{K}^*$. Finally, the optimal number of clusters $K=K^*|\underline{\alpha}_{K}^*$ is determined according to a user-defined set of clustering criteria, such as the increments in the weighted average proportion of explained inertia induced by the increase in the number of clusters and the Silhouette index \citep{Morelli2024}, the Dunn’s index, the C-index, the Calinski-Harabasz’s index, and the McClain-Rao’s index \citep{Zammarchi2024}. In this paper, we consider only the increments in the weighted average proportion of explained inertia.

\section{The role of the spatial component in hierarchical clustering}\label{sec_jointInertia}
Previously, we mentioned that a key advantage of the multi-matrix algorithm is that the choice of the weighting vector $\underline{\alpha}$ does not necessarily require the inclusion of the spatial component but rather incorporates it only when it contributes meaningfully to explaining the overall dissimilarity in the data. However, it is important to note that the weight assigned to the spatial component cannot be interpreted as a direct measure of its relevance. While it may be intuitive to think that as the spatial weight $\alpha_P$ increases, the explained inertia (and thus the proportion and normalization) also increases, it remains unclear how these quantities grow relative to each other. Indeed, if the increase in inertia were proportional to the weights of the matrices, the optimal configuration for \cite{Chavent2018} would always result in $\alpha_P = 0.5$, and the optimal outcome for \cite{Morelli2024} could be any value for $\alpha_P$.

\begin{figure}[ht!]
    \centering
    \begin{subfigure}[t]{\textwidth}
        \centering
        \begin{tabular}{p{0.1\textwidth}p{0.78\textwidth}p{0.1\textwidth}}
        &
            \includegraphics[width=0.87\linewidth]{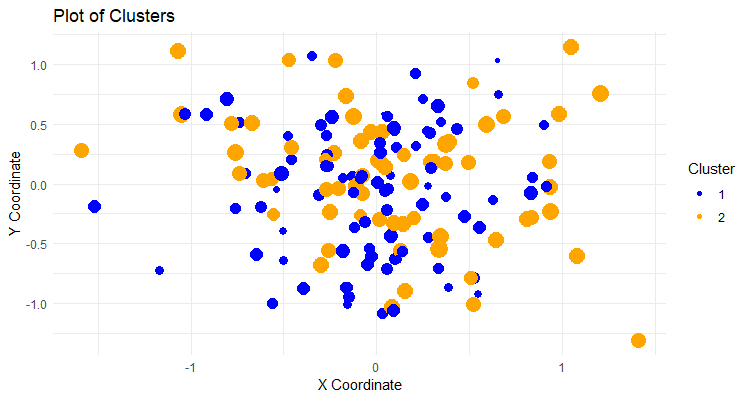} & 
            (a) 
        \end{tabular}
    \end{subfigure}   
    \begin{subfigure}[t]{\textwidth}
        \centering
        \begin{tabular}{p{0.1\textwidth}p{0.78\textwidth}p{0.1\textwidth}}
        &
            \includegraphics[width=0.87\linewidth]{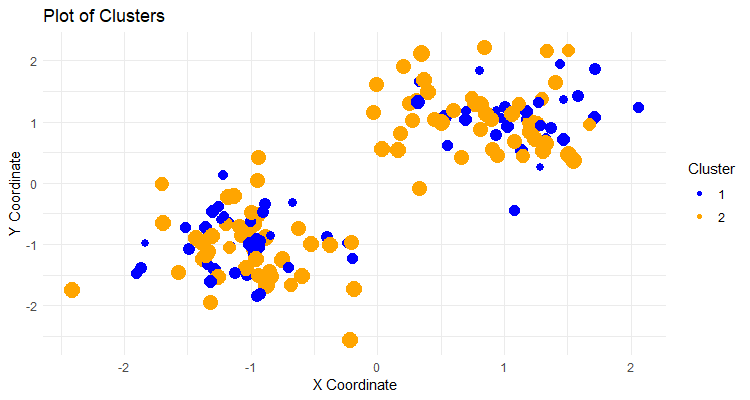} & 
            (b)
        \end{tabular}
    \end{subfigure}
    \begin{subfigure}[t]{\textwidth}
        \centering
        \begin{tabular}{p{0.1\textwidth}p{0.78\textwidth}p{0.1\textwidth}}
        &
            \includegraphics[width=0.87\linewidth]{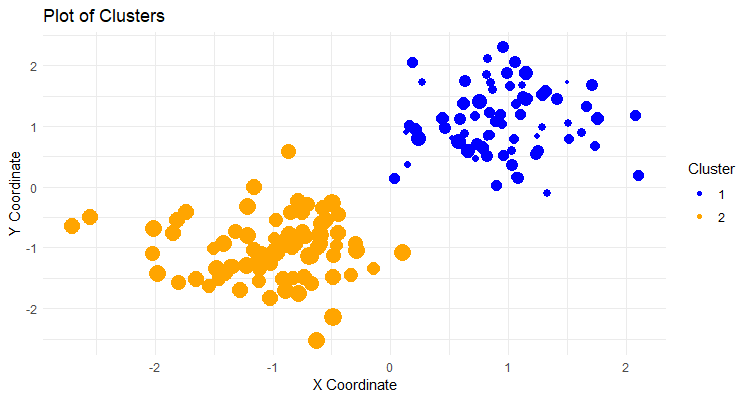} & 
            (c)
        \end{tabular}
    \end{subfigure}
    \caption{Example of clusters defined under different spatial conditions. In panel (a), the spatial component is not relevant: observations from different clusters present the same spatial distribution, and clusters are perfectly overlapping in space. In panel (b), the spatial component is relevant but not related to the phenomenon: it is possible to identify groups according to spatial distances, but this is not useful in describing the phenomenon because, in each spatial cluster, we can find observations with different features. In panel (c), the spatial component is relevant and related to the phenomenon: we can identify the same clustering partition either using spatial dissimilarities or feature dissimilarities.}
    \label{fig:clusters_cases}
\end{figure}

To illustrate this unclear behavior, we provide three simulated numerical examples (or scenarios) in which the spatial component is either (a) completely irrelevant, (b) relevant but not related to the phenomenon, or (3) relevant and useful in describing the phenomenon\footnote{Notice that the statement "the spatial component is relevant with respect to the phenomenon" refers to whether the spatial/geographical distribution of the units to be clustered is related to the phenomenon under study and thus can be useful for the purpose of adequately constructing clusters ("it is relevant for the clustering") or is unrelated to the underlying phenomenon and thus does not contribute in a useful way to the construction of clusters ("it is not relevant for the clustering").}. Without losing generalities, let us consider the case with only two dissimilarity matrices, the first containing the spatial distances and the second collecting the Euclidean distances across observations for a single feature. In Figure \ref{fig:clusters_cases}, we provide a graphical representation of the three above scenarios.

In the first case, the clusters present high dissimilarities related to the features, but they perfectly overlap with respect to the spatial coordinates. As the spatial component is not playing any role in describing the phenomenon and is not useful in maximizing the explained inertia, it will be not included in the final optimal combination of weights. In the second scenario, the partition obtained using only the features dissimilarity matrix is completely different from the one obtained using only the spatial dissimilarities. Here, the spatial component leads to the identification of groups that are well separated in space but are not related to the actual pattern of the phenomenon under study. Therefore, we expect the algorithm to return as optimal mixing parameter either $\alpha=0$ or $\alpha=1$, depending on which matrix explains a greater portion of the dissimilarity across clusters. In the third scenario, the cluster partition obtained from observing only the features is identical to the one obtained using only the spatial component. Thus, the geographical distribution is strongly related to the phenomenon of interest and will play a crucial role in building the clusters. In this latter case, the spatial component's role is clearly significant, yet any value of $\alpha$ would suffice to maximize the weighted average explained inertia because if the clustering algorithm on $D_0$ leads to the same partition obtained using only $D_1$, any combination of $D_0$ and $D_1$ will lead to the same results.

The above simulated example shows that the weight of the spatial component, represented by $\alpha$, cannot be interpreted as a measure of the actual importance of geographical information for identifying the groups of observations that characterize the patterns of the examined variables. Indeed, it is only the solution obtained from a maximization task. On the one hand, when the spatial component is included in the combined dissimilarity matrix $D$, thus $1-\alpha>0$, or $\alpha_P>0$ in multidimensional case, this indicates that spatial information is at least partially relevant in explaining the phenomenon. On the other hand, when the spatial information became more and more related to the dissimilarities across features, that is, $D_0$ and $D_1$ lead to similar clustering partitions, it is no longer necessary to include the spatial component because feature dissimilarities are able to capture a high proportion the inertia in the spatial dissimilarities and vice versa.

To properly understand the role of the spatial dissimilarity matrix in relation to the phenomenon under study, it is more appropriate to examine how the explained inertia in the resulting clusters changes when combining the two dissimilarity matrices with respect to the cases in which the features or spatial distances are used independently. The normalized proportion of explained inertia defined by \cite{Chavent2018} (Equation \ref{eq:norm_prop_inertia}) is particularly useful for this purpose because it provides the percentage of explained variability in the clustering partition obtained from the combined dissimilarity matrix with respect to the case in which the partition is obtained from one dissimilarity matrix only.

When we choose to incorporate the spatial component, we effectively renounce to a proportion of the inertia explained by the features to gain a certain amount of inertia explained within the spatial component. It follows that the actual contribution of the geographical information can be well approximated by the difference between the normalized proportion of inertia explained by the spatial component within the mixed cluster partition and the reduction in the normalized proportion of inertia explained by the features that results from including the spatial component. Specifically, in the case where there are only two matrices $D_0$ and $D_1$, we quantify the importance of spatial features through a novel indicator, namely the Joint Inertia of partition $\mathcal{P}_K^{\alpha}$, which is computed as
\begin{equation}
JI(\mathcal{P}_K^{\alpha})=\tilde{Q}_{D_1}(\mathcal{P}_K^{\alpha})-(1-\tilde{Q}_{D_0}(\mathcal{P}_K^{\alpha}))=\frac{{Q}_{D_1}(\mathcal{P}_K^{\alpha})}{{Q}_{D_1}(\mathcal{P}_K^{1})}+\frac{{Q}_{D_0}(\mathcal{P}_K^{\alpha})}{{Q}_{D_0}(\mathcal{P}_K^{0})}-1
\end{equation}

The generalization to the case of multiple dissimilarity matrices, one of which relates to the spatial component, is straightforward. Recalling the notation previously introduced, let us consider the generic partition induced by a given combination of mixing parameters (i.e., $\mathcal{P}_K^{\underline{\alpha}}$) and let $D_{-P} = [d_{p,ij}]_{p=1,\dots, P-1; i,j=1,\dots,n}$ be the dissimilarity matrices containing the distances across the $P-1$ time series for $n$ units and let $D_P = [d_{P,ij}]_{i,j=1,\dots,n}$ be the spatial dissimilarity matrix across the $n$ units, which may be computed as geodetic distances between coordinates. Let $\underline{\alpha}=[\alpha_p]_{p=1,2,\dots,P}$ be the vector of mixing weights used to compute the linear combination of dissimilarity matrices $D(\underline{\alpha})=\sum_{p=1}^P \alpha_p D_p$ such that $\sum_{p=1}^P \alpha_p = 1$. Moreover, let use define the matrix $D_{-P}(\underline{\alpha}|\alpha_P=0)=\sum_{p=1}^{P-1} \alpha_p D_p$ as the combination of the $P-1$ dissimilarity matrices excluding the spatial matrix $D_P$, and the resulting clustering partition $\mathcal{P}_K^{\alpha_{-P}}$. Similarly, define the partition obtained by using only the spatial information (i.e., by fixing $\alpha_P=1$) as $\mathcal{P}_K^{\alpha_P=1}$.

If we compute the optimal weighting vector obtained by the combination of the non-spatial dissimilarity matrices (i.e., $\underline{\alpha}^*|\alpha_P=0$), the corresponding convex combination of dissimilarity matrices (i.e., $D_{-P}(\underline{\alpha}^*|\alpha_P=0)$) will explain as much as possible the dissimilarities in non-spatial matrices, thus being $Q_{D_{-P}}(\mathcal{P}_K^{\underline{\alpha}^*|\alpha_P=0})$ greater than any other potential $Q_{D_{-P}}(\mathcal{P}_K)$ evaluated at $\underline{\alpha}|\alpha_P=0$. Considering a generic combination of $\underline{\alpha}$ and $K$, the multi-matrix Joint Inertia for the spatial dissimilarity matrix $D_P$ can be computed as
\begin{equation*}
JI_P(\mathcal{P}_K^{\underline{\alpha}})=\tilde{Q}_{D_P}(\mathcal{P}_K^{\underline{\alpha}})-(1-\tilde{Q}_{D_{-P}}(\mathcal{P}_K^{\underline{\alpha}}))=\frac{Q_{D_P}(\mathcal{P}_K^{\underline{\alpha}})}{Q_{D_P}(\mathcal{P}_K^{\alpha_P=1})}+\frac{Q_{D_{-P}}(\mathcal{P}_K^{\underline{\alpha}})}{Q_{D_{-P}}(\mathcal{P}_K^{\underline{\alpha}^*|\alpha_P=0})}-1
\end{equation*}
where $\tilde{Q}_{D_{-P}}(\mathcal{P}_K^{\underline{\alpha}})$ is the normalized proportion of explained inertia with respect to the optimal non-spatial setting and $\tilde{Q}_{D_{P}}(\mathcal{P}_K^{\underline{\alpha}})$ is the normalized proportion of explained inertia with respect to the purely-spatial setting.

As the normalized proportion of explained inertia for every dissimilarity matrix varies between $0$ and $1$, also the Joint Inertia $JI_P(\mathcal{P}_K^{\underline{\alpha}})$ will range from $0$ to $1$ (i.e., $0 \leq JI_P(\mathcal{P}_K^{\underline{\alpha}}) \leq 1$), being close to zero when the loss in the normalized proportion of explained inertia from the features dissimilarity due to the inclusion of the spatial component is similar to the normalized proportion of explained inertia in the spatial dissimilarity in the combined partition. Conversely, it will be close to $1$ when the clustering partition obtained combining features and spatial dissimilarities allows to explain a high normalized proportion of inertia, both in spatial and non-spatial dissimilarity matrices.

We stress that the Joint Inertia can be utilized to evaluate the contribution of any dissimilarity matrix relative to all other dissimilarity matrices. In this study the Joint Inertia is specifically introduced to assess the role of the spatial component, addressing our initial research objectives. However, in other contexts, the Joint Inertia of any dissimilarity matrix can be computed to gain a comprehensive understanding of which variables exhibit shared explanatory power in terms of their dissimilarities. Further technical details about the Join Inertia, both in the case of two matrices and when considering more than two matrices, are discussed in the Appendix \ref{App:joint_inertia}.

Considering the three simulated scenarios described above, in Table \ref{tab:cluster_cases} we provide the values of the normalized proportion of explained inertia in $D_0$ and $D_1$ and the Joint Inertia obtained for every case represented in Figure \ref{fig:clusters_cases}. When the spatial component is not related to the clustering partition obtained by features (e.g., scenario b), we obtain $\alpha$ being either $0$ or $1$. Thus, the normalized proportion of explained inertia in the features dissimilarity matrix (or in the spatial dissimilarity matrix) will be equal to $1$, while the normalized proportion of explained inertia in the spatial dissimilarity matrix (in the feature dissimilarity matrix) will be close to zero but strictly positive. The Joint Inertia index will be close to zero in cases (a) and (b), which is coherent with the fact that the spatial component plays no role in identifying the actual clustering patterns (conversely, well identified by the features space), while it is equal to $1$ in case (c) because the spatial information is strongly related to the underlying phenomenon and exactly describes the groups structure\footnote{It should be noted that the case $JI=1$ holds when the geographical information and the feature space describe the same group structure while being strongly related to the phenomenon under study. This happens when space and features embed the same amount of information associated with the phenomenon and resulting in the same clustering partition. So, in a sense, the geography jointly determines the features and the clusters. However, geography compresses the relevant information using only two dimensions, whereas it is reasonable to expect the same degree of information to be achieved by a much larger number of features.}. When $0<\alpha<1$, the dissimilarity matrices are combined because their combination allows us to capture an overall higher proportion of the explained inertia, thus we expect to obtain an higher value of the presented indicator. In the following Section \ref{sec_sim} we discuss the connection between the Join Inertia and the spatial information embedded in the data. In particular, we focus on how the index can be used to assess the actual relevance of the geographical dissimilarity in computing the clustering partitions.

\begin{table}[!ht]
    \centering
    \begin{tabular}{p{0.22\textwidth}|p{0.22\textwidth}|p{0.22\textwidth}|p{0.22\textwidth}}
     \centering
        \textbf{Case (a)} & \multicolumn{2}{c|}{\textbf{Case (b)}} & \textbf{Case (c)} \\ 
        Spatial component is not relevant & \multicolumn{2}{p{0.44\textwidth}|}{Spatial component is relevant but not related to the features} & Spatial component is helpful \\ \hline
        $\alpha=0$ & if $\alpha=0$ & if $\alpha=1$ & $0 < \alpha < 1$ \\ 
        $\tilde{Q}_{D_1}(\mathcal{P}_K^{\alpha})\approx0$ & $\tilde{Q}_{D_1} (\mathcal{P}_K^{\alpha})\approx0$ & $\tilde{Q}_{D_1}(\mathcal{P}_K^{\alpha})=1$ & $\tilde{Q}_{D_1}(\mathcal{P}_K^{\alpha})=1$ \\
        $\tilde{Q}_{D_0}(\mathcal{P}_K^{\alpha})=1$  & $\tilde{Q}_{D_0}(\mathcal{P}_K^{\alpha})=1$ & $\tilde{Q}_{D_0}(\mathcal{P}_K^{\alpha})\approx0$ & $\tilde{Q}_{D_0}(\mathcal{P}_K^{\alpha})=1$ \\
        $JI \approx 0$ & $JI \approx 0$ & $JI \approx 0$ & $JI=1$
    \end{tabular}
    \caption{Choice of $\alpha$, normalized proportion of explained inertia and resulting Joint Inertia in different examples of clusters under different spatial conditions. Cases are represented in Figure \ref{fig:clusters_cases}.}
    \label{tab:cluster_cases}
\end{table}

\section{Simulation experiment}\label{sec_sim}
In this section, we provide a simulation experiment aiming to compare the performance of the clustering algorithm based on both \cite{Morelli2024} and \cite{Chavent2018} variants in choosing the mixing parameter $\alpha$ when considering several settings for the actual contribution of the geographical information. In particular, we provide insights into the role of the spatial component by using an overlapping parameter that controls the distance across the means of the spatial distribution of the true clusters. Under the simulation design, we compare the algorithms in terms of accuracy, precision, sensitivity \citep{NaiduEtAl2023}, and the adjusted rand index \citep{Gordon1999} for multi-class data. 

Using a Monte Carlo simulation scheme with $N=500$ repetitions, we simulate samples of size $n=100$ observations from a mixture of three-dimensional Gaussian distribution in which the first dimension refers to a unique non-spatial feature and the other two dimensions refer to the spatial coordinates. We simulate data clustered into $K=4$ different groups. We indicate with $\pi_1,..,\pi_k$ the probability to sample an observation belonging to each cluster $k=1,\dots ,4$ and, according to the cluster, we simulate the variable $Z$ and the spatial coordinates $(X,Y)$. Specifically, we define the distribution of the variable as $Z_k \sim N(\mu_k,\sigma)$, where $\mu_k=(\mu_1,\dots,\mu_4)$ and $\sigma=v$, and the spatial component $(X,Y)_{k} \sim N(\mu_{sp,k}, \Sigma_{sp})$. The means of the group-specific spatial coordinates are defined according to an overlapping parameter, namely $d$, which can vary between $0$ and $1$, and it determines the proximity across the spatial means of the clusters. In particular, $\mu_{sp,k}=(\mu_{sp,1},\mu_{sp,2},\mu_{sp,3},\mu_{sp,4})=((d,d),(-d,d),(d,-d),(-d,-d))$ while $\Sigma_{sp}=v_{sp}\mathbb{I}_{2}$. Notice that, given this simulation setting, as the overlapping parameter grows from $0$ to $1$, the centroids of the groups tend to diverge, thus generating well-distanced clusters and enhancing the relevance of the geographical information in clustering the units.

\begin{figure}[ht]
    \centering
    \begin{tabular}{cc} 
        \begin{minipage}{0.8\textwidth}
            \centering
            (a)\includegraphics[width=0.90\linewidth]{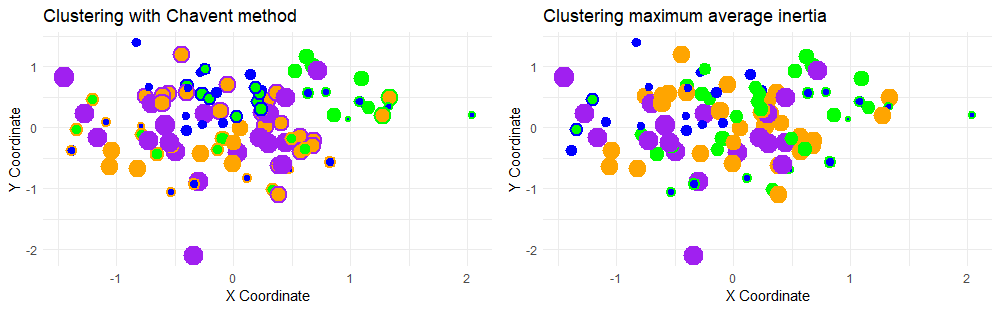}\\
            (b)\includegraphics[width=0.90\linewidth]{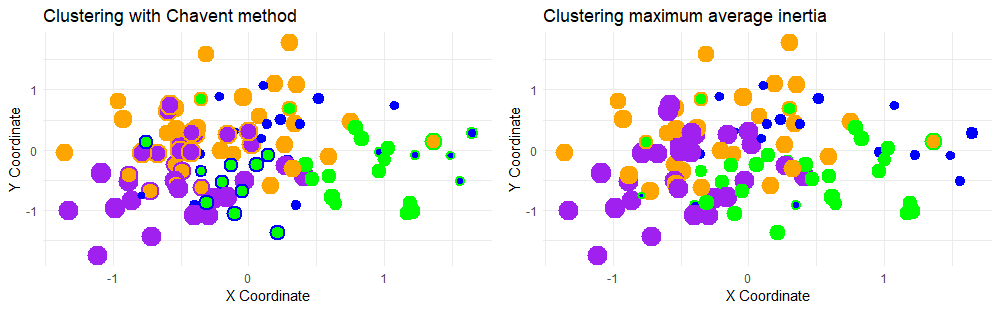}\\
            (c)\includegraphics[width=0.90\linewidth]{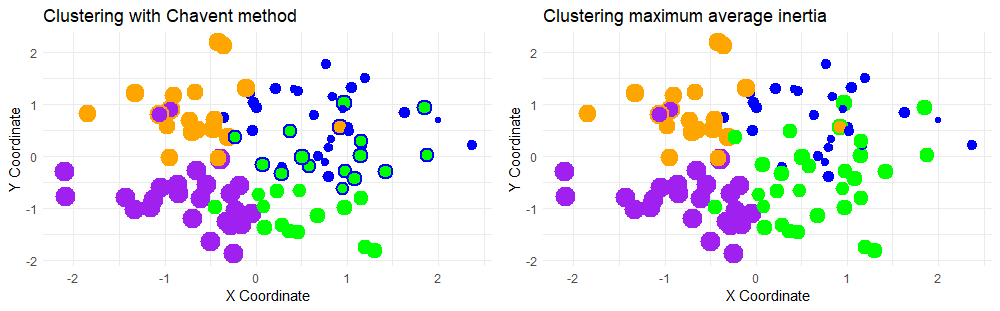}\\
            (d)\includegraphics[width=0.90\linewidth]{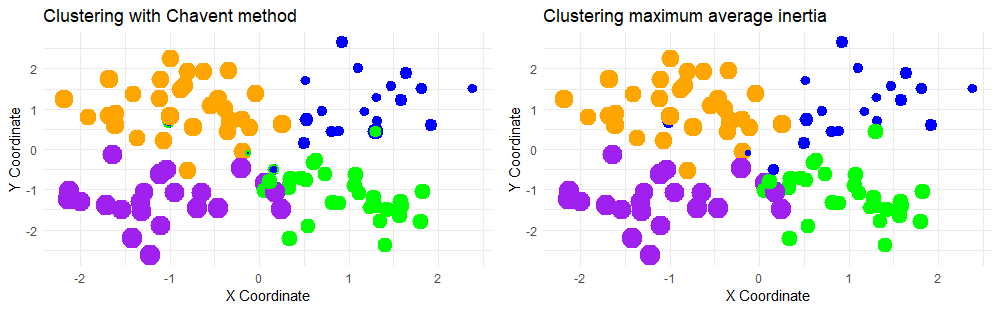}
        \end{minipage}
        &
        \begin{minipage}{0.18\textwidth}
            \centering
            \includegraphics[width=1\linewidth]{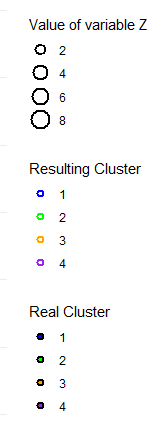}
        \end{minipage}
    \end{tabular}
    \caption{Clustering partition in four simulations with different overlapping parameters $d \in \{0,1/3,2/3,1 \}$ respectively in panel (a), (b), (c), and (d). The color of the points refers to the true cluster from which the observation is drawn, while the color of the border refers to the resulting clustering partition obtained by applying the algorithm from \cite{Chavent2018} (left column) and our methodology from \cite{Morelli2024} (right column).}
    \label{fig:sim_plot}
\end{figure}

To exemplify the contribution of the overlapping parameter $d$, in Figure \ref{fig:sim_plot}, we depict some examples of the clustering partition that can be obtained for different values of the overlapping parameter and distinguishing among \cite{Chavent2018} (left column) and \cite{Morelli2024} (right column) criteria. The plots are simulated by setting $\pi_k=1/K$, $\mu_k=(2,4,6,8)$, $v=0.4$, $v_{sp}=0.4$, and considering the following set of values for the overlapping parameter $d \in \{0,1/3,2/3,1\}$ (i.e., sorted top-down).

From the figure, it is possible to observe that the maximum average criterion is able to detect a clustering partition similar to the real one, even when the spatial component is not relevant or less relevant, that is, in panel \textit{a} and \textit{b}. When the distance in the spatial location of the cluster is higher (e.g., in panel \textit{c} and \textit{d}), clustering performance improves using both methodologies. Results reported in Table \ref{tab:sim} confirm the previous statements. Indeed, while \cite{Chavent2018} methodology allows identifying a reasonable clustering partition, the \cite{Morelli2024} methodology demonstrates improved clustering performance in terms of accuracy and adjusted rand index. This enhancement is particularly notable in scenarios where the spatial component plays a limited role, such as when clusters fully or partially overlap in space.

\begin{table}[!ht]
    \centering
    \begin{tabular}{c|ccc|ccc|c}
 
        ~ & \multicolumn{3}{|c|}{\cite{Chavent2018}} & \multicolumn{3}{|c|}{\cite{Morelli2024}} & ~  \\ 
        $d$ & $\alpha$ & Accuracy & Adj. RI & $\alpha$ & Accuracy & Adj. RI & Joint Inertia  \\ \hline
        0 & 0.75 & 0.43 & 0.27 & 0.35 & 0.87 & 0.74 & 0.17 \\ 
        1/3 & 0.65 & 0.63 & 0.34 & 0.50 & 0.93 & 0.83 & 0.49 \\ 
        2/3 & 0.85 & 0.82 & 0.61 & 0.45 & 0.97 & 0.93 & 0.86 \\ 
        1 & 0.75 & 0.96 & 0.91 & 0 & 1.00 & 1.00 & 0.93 
    \end{tabular}
            \caption{Comparison of the performance resulting from clustering partition according to \cite{Chavent2018} and \cite{Morelli2024} algorithm and the resulting Joint Inertia to assess the role of the spatial component in the partition.}
        \label{tab:sim}
\end{table}

Moreover, we confirm that the parameter $\alpha$ can not be interpreted as a measure of assessment of the role of the spatial component as the optimized values do not reflect the degree of overlapping of the data-generating process. On the other hand, the Joint Inertia index appears effective in assessing the relevance of the spatial component in the resulting cluster distribution.

Below, we present the results from the Monte Carlo experiment and random assignment for parameters $\pi_k \sim U (0.15,0.35)$, $v \sim U(0.1,0.6)$, $v_{sp}=0.4$ and $d \sim U(0,1)$.
From Table \ref{tab:sim_mc}, we can clearly observe that maximum average criterion by \cite{Morelli2024} is able to find clustering partitions that are very similar to the true partitions and performs remarkably better in identifying the actual mixing parameter $\alpha$ with respect to the \cite{Chavent2018} criterion.

\begin{table}[!ht]
    \centering
    \begin{tabular}{c|cc|cc|cc|cc}
        & \multicolumn{2}{|c|}{\textbf{Accuracy}} & \multicolumn{2}{|c|}{\textbf{Precision}} & \multicolumn{2}{|c|}{\textbf{Sensitivity}} & \multicolumn{2}{|c}{\textbf{Adj. Rand Index}}  \\ 
        & Chavent & Morelli  & Chavent & Morelli & Chavent & Morelli & Chavent & Morelli  \\ \hline
        Mean & 0.74 & 0.85 & 0.75 & 0.85 & 0.74 & 0.84 & 0.55 & 0.71  \\ 
         Median & 0.76 & 0.91 & 0.78 & 0.91 & 0.76 & 0.90 & 0.52 & 0.77  \\ 
          sd & 0.16 & 0.14 & 0.17 & 0.15 & 0.16 & 0.15 & 0.23 & 0.22  \\  
    \end{tabular}
            \caption{Comparison of the performance resulting from clustering partition according to \cite{Chavent2018} and \cite{Morelli2024} algorithm in Monte Carlo simulation, in terms of Accuracy, Precision, Sensitivity ad Adjusted Rand Index.}
        \label{tab:sim_mc}
\end{table}

\begin{figure}
    \centering
    \includegraphics[width=1\linewidth]{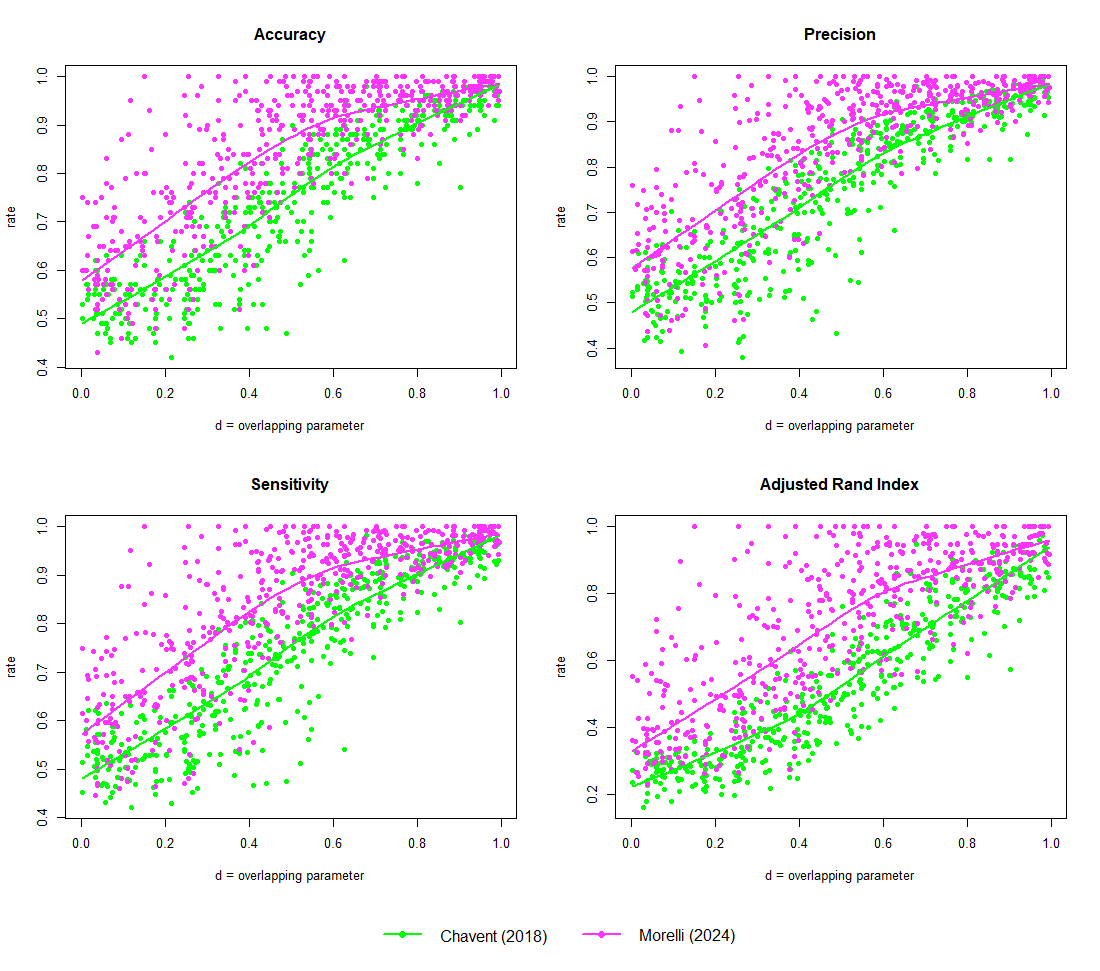}
    \caption{Clustering performance comparison between \cite{Chavent2018} and \cite{Morelli2024} in terms of Accuracy, Sensitivity, Precision, and Adjusted Rand Index, with respect to the overlapping parameter $d$.}
    \label{fig:accuracy}
\end{figure}

In Figure \ref{fig:accuracy} we provide a comparison of the resulting clustering performance with respect to the overlapping parameter $d$ used in the simulation. It is evident that \cite{Morelli2024} algorithm performs generally better, in particular when the overlapping parameter is low when clusters are partially overlapping in space. As expected, when the overlapping parameter $d$ is close to $1$, both algorithms are able to find exactly the true partition because, as explained in the previous section, when the feature dissimilarity and the spatial dissimilarity lead to the same partition, any combination of them will lead to it.

\begin{figure}
    \centering
    \includegraphics[width=1\linewidth]{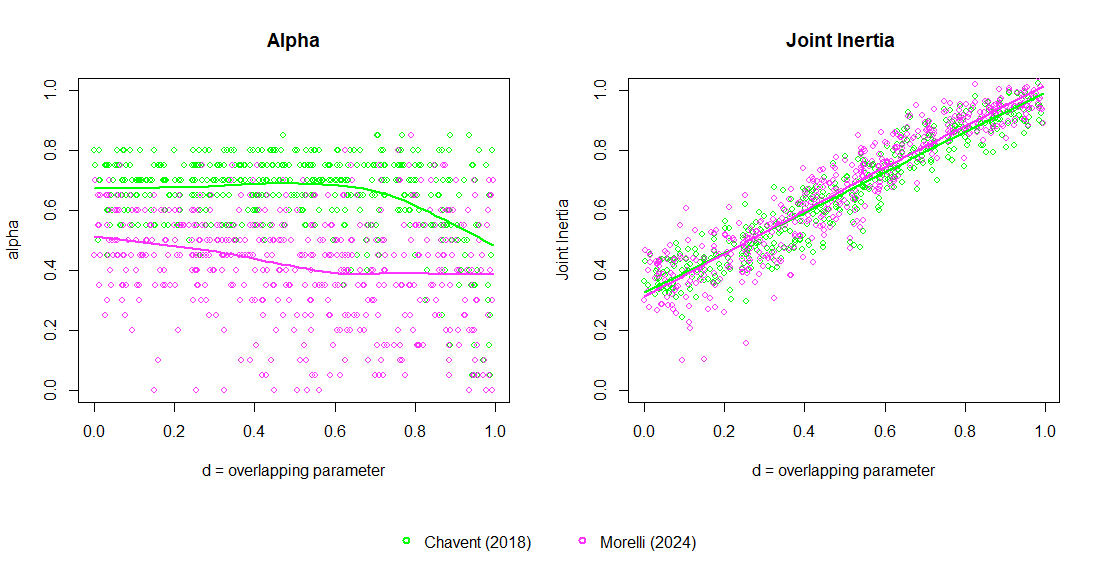}
    \caption{In panel (a), on the left, we can observe the resulting $\alpha$ according to \cite{Chavent2018} and \cite{Morelli2024} when considering clusters with different overlapping degrees (i.e., when the overlapping parameter $d$ varies between $0$ and $1$). In panel (b), on the right, we report the resulting Joint Inertia obtained in the simulation experiment with respect to the overlapping parameter $d$.}
    \label{fig:alpha_inertia}
\end{figure}

In Figure \ref{fig:alpha_inertia}, we provide several insights into the relationship between the spatial distance among centroids, determined by the overlapping parameter $d$, the mixing parameter $\alpha$, and the Joint Inertia index. In the left panel, as the overlapping parameter grows (from left to right), the mixing parameter selected by both methods seems to be unaffected, confirming that the weight assigned to the spatial dissimilarity matrix does not provide information about the actual contribution of the spatial component in determining the clusters. Conversely, the right panel remarkably supports the use of the Joint Inertia index in quantifying the importance of the spatial component. Indeed, a strong and positive correlation between the overlapping parameter $d$ and the Joint Inertia is detected, both using $\alpha_{chavent}$ and $\alpha_{morelli}$. The positive correlation is definitely consistent with the interpretation of the overlapping parameter $d$ given above, that is, as $d$ grows from $0$ to $1$, the distance between the centroids of the groups tends to increase, increasing the informative power induced by the spatial/geographical compared to the non-spatial distance, exactly as described by the Joint Inertia.

\section{Dynamic and Regional Clustering of GHGs Emissions}\label{sec_app}

In this section, we present an application of spatiotemporal clustering on environmental data concerning the annual GHGs emissions at the regional level for Europe. Our aim is to explore regional patterns in the environmental impact caused by human activities on air quality, which may also depend on the geographical and morphological characteristics of the area as well as on the productive specialization of the regions. 


We analyze the annual average emissions per $\text{km}^2$ of several GHGs, that is,  methane, nitrous oxide, fluorinated gases (F-gases), and carbon dioxide produced by the main economic sectors, namely agriculture, buildings, energy, industry, transport, and waste. Data are measured in tons per $\text{km}^2$ of $\text{CO}_2$ equivalent.

Greenhouse gases in Europe originate from different activities, each of them contributing to climate change in unique ways \citep{GHGsJRC}. Methane ($\text{CH}_4$) emissions primarily arise from agriculture, especially livestock farming and waste management, as well as fossil fuel extraction and processing. Nitrous oxide ($N_2O$) is largely emitted through agricultural practices, such as the use of nitrogen-based fertilizers, with smaller contributions from industrial processes and fossil fuel combustion. F-gases, synthetic in nature, are used in refrigeration, air conditioning, and industrial processes like aluminum production. Carbon dioxide ($CO_2$), the most abundant GHGs, is predominantly released through fossil fuel combustion for energy, transportation, and industrial activities, as well as deforestation and cement production.

We believe that the identification of patterns and dynamics in regional GHGs emissions within a spatiotemporal framework is a valuable contribution that extends the current knowledge on the status of environmental degradation in Europe, enhancing the awareness of citizens and institutions regarding the policies to be planned to contrast climate change and global warming.

\subsection{Dataset description}
\begin{figure}
    \centering
    \includegraphics[width=1\linewidth]{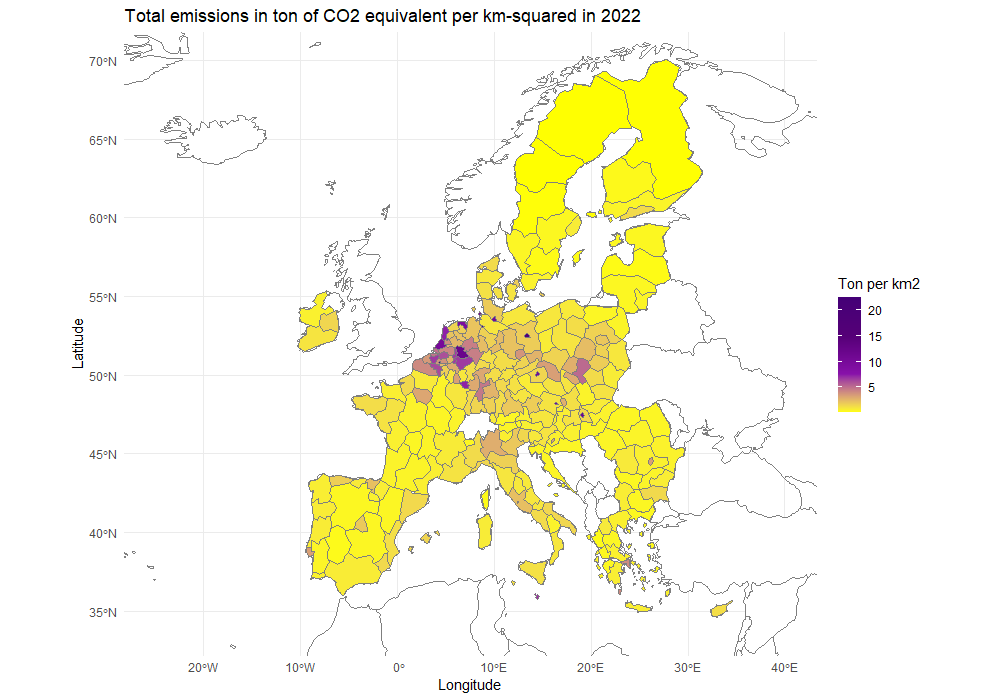}
    \caption{Map of total emission per $\text{km}^2$ in 2022 in each region at NUTS2 level.}
    \label{fig:map_emission22}
\end{figure}

The data of interest relate to atmospheric emissions of sector-specific greenhouse gases measured at the regional level provided by the Emissions Database for Global Atmospheric Research (EDGAR) database \citep{GHGsJRC} within the Annual Regional Database of the European Commission (ARDECO) \citep{ARDECO} project. In particular, we consider yearly measurements of GHGs emissions from 1990 to 2022 (i.e., $T=33$ time stamps) for the $N=234$ regions\footnote{Notice that we excluded regions classified as extra-continental territories and those located in non-European continents, ensuring that the extremely high distances involving these regions do not render the spatial dissimilarities among the remaining regions insignificant.} in the European Union, in 27 countries, identified according to the Nomenclature of territorial units for statistics - Level 2 (NUTS-2) classification \citep{NUTS}. We also collect the area of NUTS-2 regions in 2022 from Eurostat \footnote{https://ec.europa.eu/eurostat/databrowser/view/REG\_AREA3/default/table} dataset in order to obtain the annual emissions in tons per $\text{km}^2$ of $CO2$ equivalent, by gases and by sector. In Figure \ref{fig:map_emission22} we represent a map of the total GHGs emissions per $\text{km}^2$ in 2022 at the regional level. 

We run the spatiotemporal cluster analysis twice, considering two different datasets. In the first case, we implement the clustering by aggregating GHGs emissions according to the type of gas (i.e., by summing the emissions at the sectoral level for each gas). In the second case, we aggregate the emissions with respect to the sectors, thus by summing the gas-specific emissions for each sector. In both cases, we utilize the Dynamic Time Warping (DTW) strategy as the chosen distance metric. The Dynamic Programming approach using a warping function has been introduced by \cite{Sakoe1971ADP,Sakoe197843} in the spoken word recognition field as a time-normalization algorithm. \cite{Berndt1994} have implemented DTW distance to detect patterns in time series. We compute the spatial dissimilarity matrix as the geodetic distance across the coordinates of the centroids of each region.

\subsection{Spatiotemporal clusters for emissions by gas}\label{sec:spt_gas}
\begin{figure}
    \centering
    \includegraphics[width=0.8\linewidth]{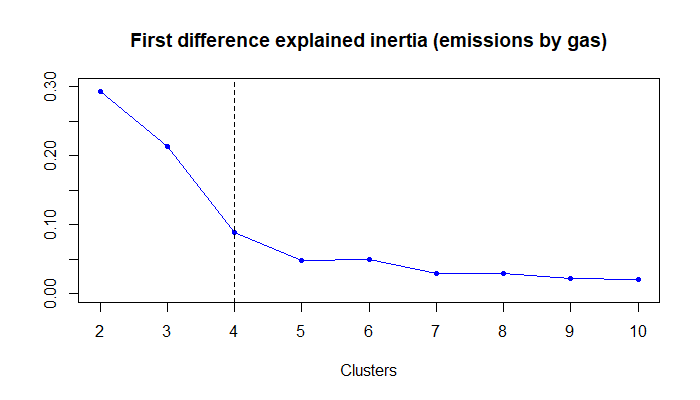}
    \caption{Increment in the weighted average proportion of explained inertia generated by a unitary increase in the number of clusters. Recall that, for each $K$, we considered the best combination of the dissimilarity matrices according to the $\alpha^*_K$ found using the maximization inertia criterion.}
    \label{fig:diff_inertia_gas}
\end{figure}

We run the algorithm described in \cite{Morelli2024} setting $\Delta \alpha=0.05$, for $K=2,\dots,10$, combining $P=5$ dissimilarity matrices referring to the $CH_4$, F-Gases, Fossil-$CO_2$, $N_2O$ emission time series and the spatial component. We select $K^*=4$ according to the elbow rule, observing the weighted average proportion of explained inertia, which can be interpreted as the gain in explained variability due to an additional cluster. In Figure \ref{fig:diff_inertia_gas} we represent the first difference in the weighted average proportion of explained inertia, highlighting our choice of $K^*=4$.

Consequently, we selected $\alpha_{4,p}^*=(0.20, 0.15, 0.10, 0.20, 0.35)$ as the optimal weighting vector for the dissimilarity matrices. Table \ref{tab:spt_gases} provides a summary of the weights $\alpha^*_p$, the explained inertia (both proportion and normalized), and the Joint Inertia associated with the resulting clustering partition for each dissimilarity matrix included in the algorithm.

From Table \ref{tab:spt_gases}, it is possible to observe that the partition obtained with $K^*=4$ clusters allows us to explain almost $40\%$ of the DTW dissimilarities across $CO_2$ time series and about $60\%$ of the dissimilarities across time series of $CH_4$, F-gases, $N_2O$ and the spatial dissimilarity. The latter has a Joint Inertia of $0.55$, which allows us to conclude that spatial information is crucial when determining the clusters composition, and therefore, the spatial distance is useful in the identification of patterns to explain well the data under consideration. In particular, through the resulting clustering partition $\mathcal{P}_K^{\underline{\alpha}^*}$ we are able to incorporate $80\%$ of the explained inertia in the spatial component, relative to the case in which we consider only $D_{sp}$ and, considering the non-spatial dissimilarities, we are still able to include $75\%$ of the explained inertia with respect to the partition obtained excluding the spatial component.
All the dissimilarity matrices have a weight greater than zero and they all seem to be relevant in the composition of the clusters, in fact they have a rather high Joint Inertia. it is possible to note that the weights assigned to the dissimilarity matrices do not always reflect the relevance of the matrix itself in relation to the clustering partition obtained, for example the matrix $D_{CO_2}$ is assigned a weight of $0.2$, greater than the $0.15$ and $0.10$ weights assigned to matrices $D_{\text{F-gases}}$ and $D_{N_2O}$, yet the proportion of inertia explained in the first,which is $0.38$, is less than the inertia explained in the other two, respectively $0.68$ and $0.62$.

\begin{table}[!ht]
    \centering
    \resizebox{0.8\textwidth}{!}{%
    \begin{tabular}{c|ccccc}
         & $D_{CH_4}$ & $D_{\text{F-gases}}$ & $D_{N_2O}$ & $D_{CO_2}$ & $D_{sp}$   \\ \hline
        $\alpha^*_p$ & 0.20 & 0.15 & 0.10 & 0.20 & 0.35  \\ 
        $\mathcal{Q}_{D_p}(\mathcal{P}_K^{\underline{\alpha}})$ & 0.67 & 0.68 & 0.62 & 0.38 & 0.58  \\ 
         $\tilde{\mathcal{Q}}_{D_p}(\mathcal{P}_K^{\underline{\alpha}})$ & 0.75 & 0.82 & 0.66 & 0.46 & 0.80  \\
          $\tilde{\mathcal{Q}}_{D_{-p}}(\mathcal{P}_K^{\underline{\alpha}})$ & 0.94 & 0.93 & 1.03 & 0.94 & 0.75  \\
         $JI_p(\mathcal{P}_K^{\underline{\alpha}^*})$ & 0.69 & 0.75 & 0.68 & 0.41 &  0.55 \\ 
    \end{tabular}%
    }
    \caption{Summary of the $\underline{\alpha}^*$ weights and the inertia (proportion and normalized) returned by the spatiotemporal clustering of emissions by gases at the optimal solution $K^*=4$. 
    $\mathcal{Q}_{D_p}(\mathcal{P}_K^{\underline{\alpha}})$ is the proportion of explained inertia in $D_p$; $\tilde{\mathcal{Q}}_{D_p}(\mathcal{P}_K^{\underline{\alpha}})$ is the normalized proportion of inertia in $D_p$; $\tilde{\mathcal{Q}}_{D_{-p}}(\mathcal{P}_K^{\underline{\alpha}})$ is the normalized proportion of inertia excluding $D_p$; and $JI_p(\mathcal{P}_K^{\underline{\alpha}^*})$ is the Joint Inertia relative to $D_p$.}
    \label{tab:spt_gases}
\end{table}

\begin{figure}
    \centering
    \includegraphics[width=1\linewidth]{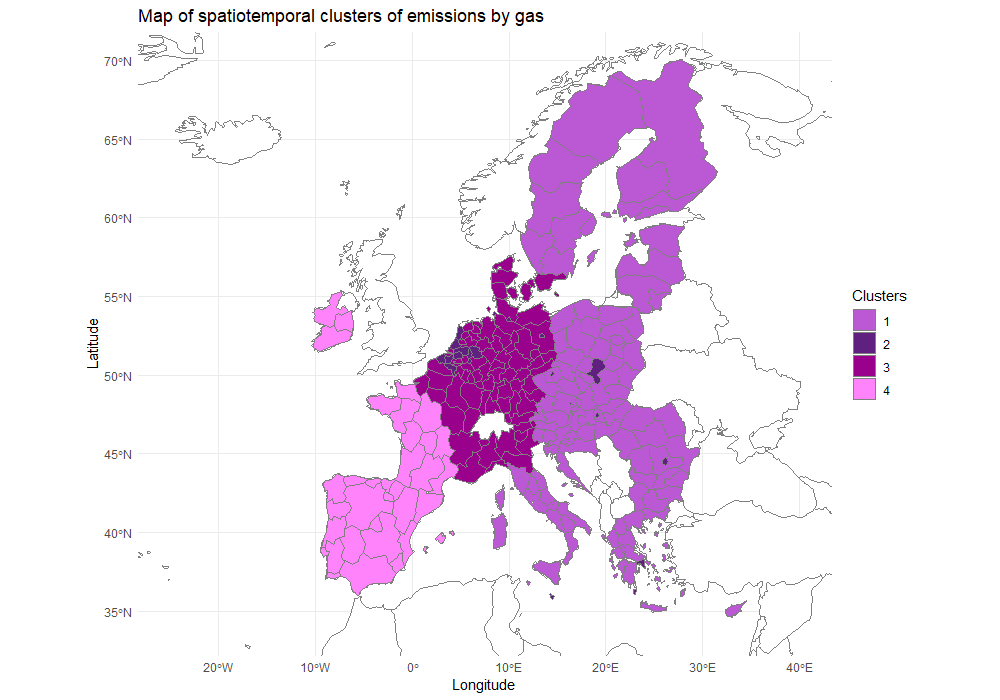}
    \caption{Map of spatiotemporal clusters obtained combining the dissimilarity matrices of the time series of emissions of different gases between 1990 and 2022, and the geodetic distances across centroids of NUTS2 level regions.}
    \label{fig:map_sp_clu_gas}
\end{figure}

\begin{figure}
    \centering
    \includegraphics[width=1\linewidth]{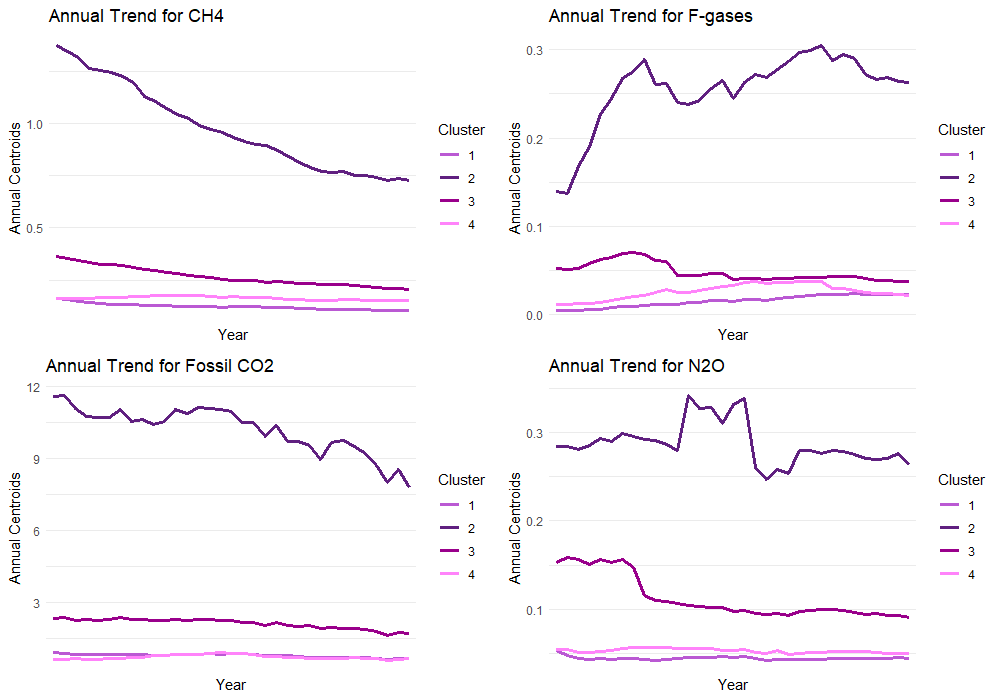}
    \caption{Annual average of emission per $\text{km}^2$ of gases for spatiotemporal clusters obtained combining the dissimilarity matrices of the time series of emissions of different gases between 1990 and 2022, and the geodetic distances across centroids of NUTS2 level regions.}
    \label{fig:sp_clu_centroids_gas}
\end{figure}

In Figure \ref{fig:map_sp_clu_gas} we present a map of the regions belonging to different clusters and in Figure \ref{fig:sp_clu_centroids_gas} we report the average annual emission in each cluster by gas. 

Cluster 1 includes regions located in East Europe, from the Scandinavian peninsula to the Balkan countries, the central and southern areas of Italy and the eastern area of Austria. It shows the lowest level of $CH_4$, F-gases, $CO_2$ and $N_2O$ emissions compared to the other clusters and it appears to be quite stable over time, except for F-gases emissions that are slowly increasing.   

Regions of the Netherlands, Malta, and several metropolitan areas in East-Central Europe belong to Cluster 2, which is characterized by emission levels that are remarkably higher—by a significant margin—than those observed in any other cluster. As for the dynamics of $CH_4$ emissions, these exhibit a decreasing trend, nearly halving between 1990 and 2022. Similarly, $CO_2$ emissions show a decline over the years, dropping from approximately 12 tons per $\text{km}^2$ to about 8 tons per $\text{km}^2$. In contrast, emissions of $F$-gases experience a sharp increase during the 1990s, stabilizing thereafter at a level of approximately 0.25 tons per $\text{km}^2$. Meanwhile, emissions of $NO_2$ display fluctuations over time without any discernible trend.


Moving to Cluster 3, we can see from the map that it incorporates regions from North-central Europe, mainly Denmark, Germany, east part of France, and Northen Italy. It appears to be the second worst cluster in term of emission per $\text{km}^2$, showing a slowly decreasing trend in $CH_4$ emissions, quite stable level of $CO_2$ and F-gases emissions and a drop in $NO_2$ emissions in the last '90s.

Finally, cluster 4 consists of regions from Ireland, Spain, Portugal, and the West part of France. Regions in this cluster exhibit very low levels of emissions, comparable to those in Cluster 1. These emissions are characterized by a slightly increasing trend for F-gases and stable levels for $CH_4$, $CO_2$, and $N_2O$.

In order to better examine the role of the spatial component on cluster formation, we performed non-spatial clustering as a robustness check. The clusters exhibit greater overlap compared to the results obtained with spatiotemporal clustering. However, it remains possible to distinguish regions in Central-Northern Europe as those with the highest levels of emissions per $\text{km}^2$. A full discussion on the two robustness checks and the corresponding results is reported in Appendix \ref{App:ts_clu_gas}.

\subsection{Spatiotemporal clusters for emissions by sector}\label{sec:spt_sector}
\begin{figure}
    \centering
    \includegraphics[width=0.8\linewidth]{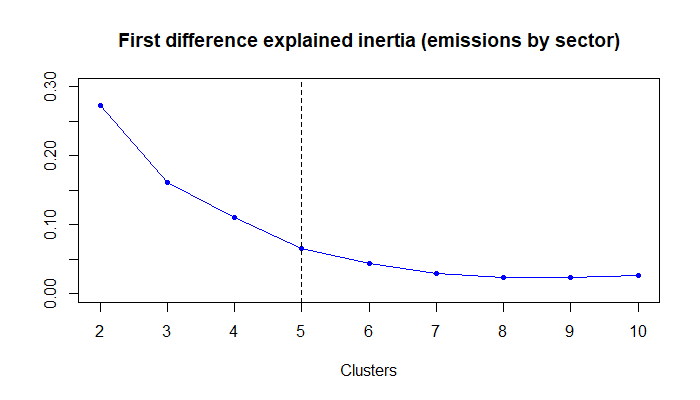}
    \caption{Increment in the weighted average proportion of explained inertia generated by a unitary increase in the number of clusters. Recall that, for each $K$, we considered the best combination of the dissimilarity matrices according to the $\alpha^*_K$ found using the maximization inertia criterion.}
    \label{fig:diff_inertia_sector}
\end{figure}

The second spatiotemporal clustering considers the GHGs emissions by sector. We compute the dissimilarity matrices of the total GHGs emissions in agricultural, buildings, energy, industry, transport, and waste sectors, and we combine them with the spatial dissimilarity matrix, thus having $P=7$. We set $\Delta \alpha=0.1$, for $K=2,\dots,10$, and we select $K^*=5$ according to the elbow rule and $\alpha_{5,p}^*=(0.20, 0.10, 0.10, 0.20, 0, 0.10, 0.30)$ as the weight of the dissimilarity matrices. In Figure \ref{fig:diff_inertia_sector}, we represent the first difference in the weighted average proportion of explained inertia, highlighting our choice of $K^*=5$.

In Table \ref{tab:spt_sector}, we report a summary of the weights $\underline{\alpha}^*$, the explained inertia (proportion and normalized), and the Joint Inertia in the resulting clustering partition for each dissimilarity matrix included in the algorithm. 

From Table \ref{tab:spt_sector}, it is possible to observe that the partition obtained with $K^*=5$ clusters allows to explain at least $35\%$ of the proportion of inertia in each dissimilarity matrix and the $36\%$ of the inertia with respect to the cases in which clusters are computed only relying on one dissimilarity matrix. Each dissimilarity matrix seems to be quite helpful in obtaining the resulting partition. In particular, the spatial component presents a Joint Inertia equal to $0.65$, and it incorporates $0.83$ of the inertia normalized to the explained inertia in the partition obtained from the purely spatial case. On the other hand, including the spatial component, we are still able to capture $0.73$ of the inertia in the non-spatial component.
Although all the dissimilarity matrix $D_\text{Tran}$ has weights equal to $0$, thus it is not included in the combined dissimilarity matrix, the resulting partition incorporates $0.61$ of the inertia, which corresponds to the $67\%$ normalized to the explained inertia from the partition obtain only considering $D_\text{Tran}$.

\begin{table}[!ht]
    \centering
    \resizebox{0.95\textwidth}{!}{%
    \begin{tabular}{c|ccccccc}
         & $D_{\text{Agri}}$ & $D_{\text{Build}}$ & $D_{\text{Energy}}$ & $D_{\text{Ind}}$ & $D_{\text{Tran}}$ & $D_{\text{Waste}}$ & $D_{\text{sp}}$  \\ \hline
       $\alpha^*_p$ & 0.20 & 0.10 & 0.10 & 0.20 & 0.00 & 0.10 & 0.30  \\ 
         $\mathcal{Q}_{D_p}(\mathcal{P}_K^{\underline{\alpha}})$  & 0.69 & 0.57 & 0.35 & 0.58 & 0.62 & 0.61 & 0.65 \\ 
        $\tilde{\mathcal{Q}}_{D_p}(\mathcal{P}_K^{\underline{\alpha}})$ & 0.72 & 0.59 & 0.36 & 0.60 & 0.67 & 0.70 & 0.83  \\ 
        $\tilde{\mathcal{Q}}_{D_{-p}}(\mathcal{P}_K^{\underline{\alpha}})$ & 0.93 & 0.99 & 1.00 & 0.95 & 0.99 & 0.94 & 0.73  \\ 
        $JI_p(\mathcal{P}_K^{\underline{\alpha}^*})$ & 0.65 & 0.58 & 0.35 & 0.55 & 0.67 & 0.64 & 0.56 \\ 
    \end{tabular}%
    }
    \caption{Summary of the $\underline{\alpha}^*$ weights and the inertia (absolute, relative, and normalized) returned by the spatiotemporal clustering of emissions by gases at the optimal solution $K^*=4$. 
    $\mathcal{Q}_{D_p}(\mathcal{P}_K^{\underline{\alpha}})$ is the proportion of explained inertia in $D_p$; $\tilde{\mathcal{Q}}_{D_p}(\mathcal{P}_K^{\underline{\alpha}})$ is the normalized proportion of inertia in $D_p$; $\tilde{\mathcal{Q}}_{D_{-p}}(\mathcal{P}_K^{\underline{\alpha}})$ is the normalized proportion of inertia excluding $D_p$; and $JI_p(\mathcal{P}_K^{\underline{\alpha}^*})$ is the Joint Inertia relative to $D_p$.}
    \label{tab:spt_sector}
\end{table}

\begin{figure}
    \centering
    \includegraphics[width=1\linewidth]{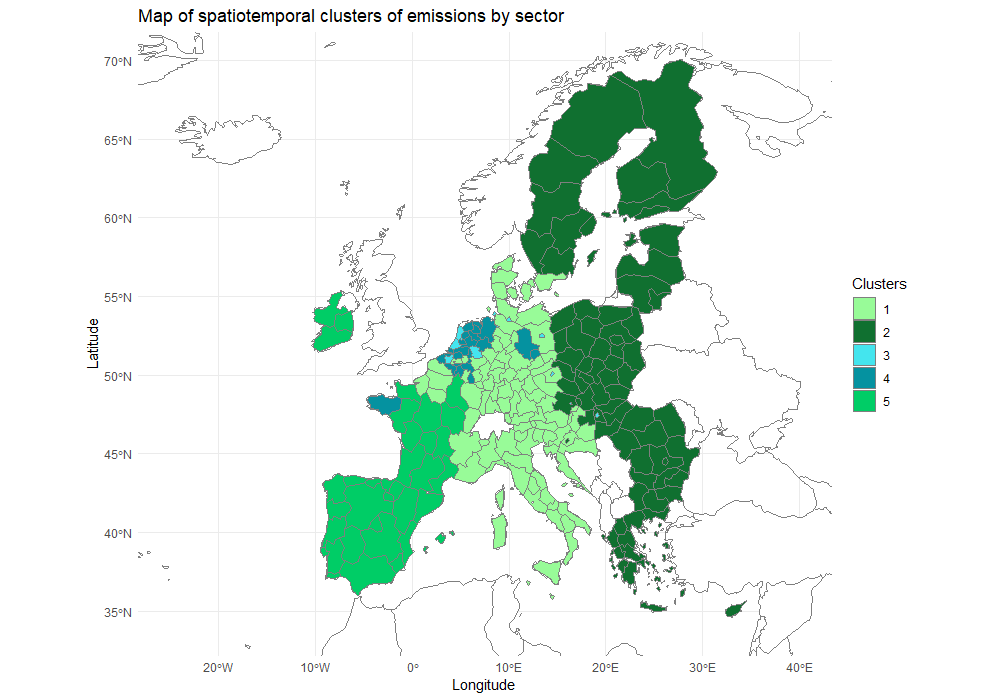}
    \caption{Map of spatiotemporal clusters obtained combining the dissimilarity matrices of the time series of emissions of different sectors between 1990 and 2022, and the geodetic distances across centroids of nuts2 level regions.}
    \label{fig:map_sp_clu_sect}
\end{figure}

\begin{figure}
    \centering
    \includegraphics[width=1\linewidth]{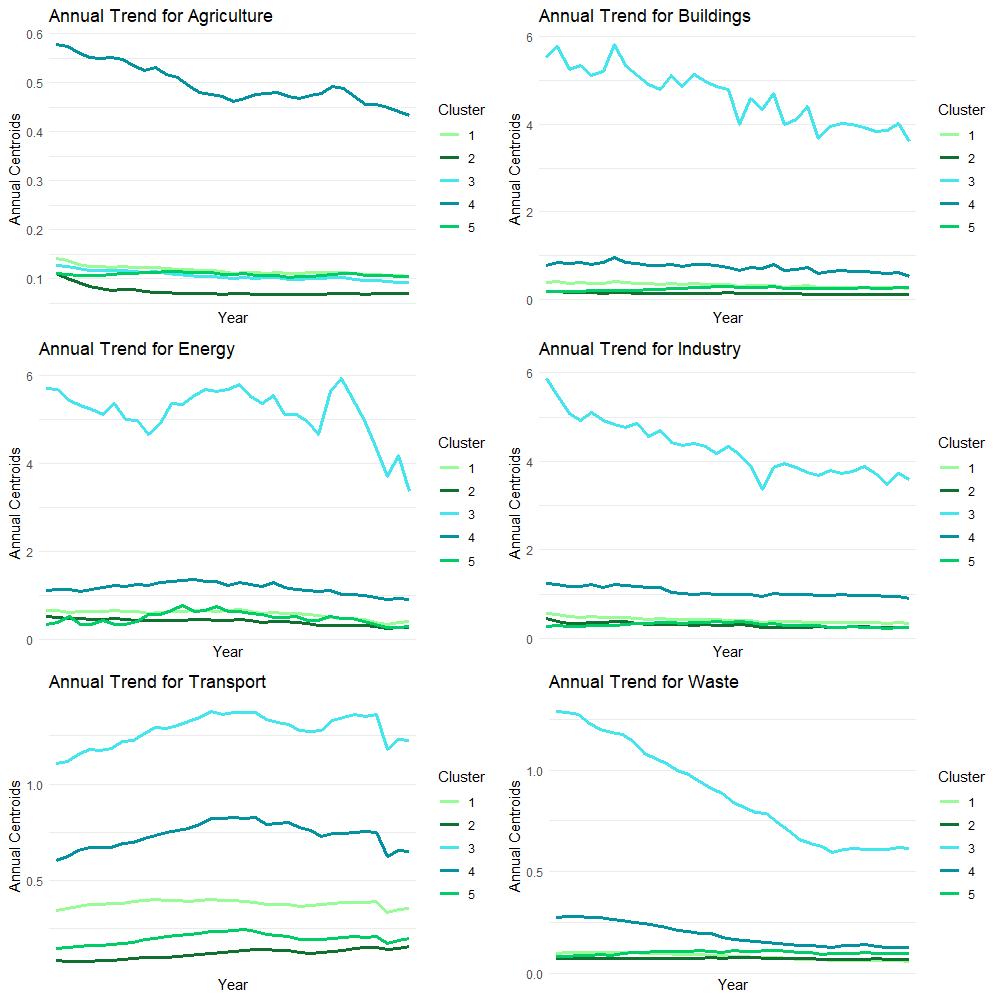}
    \caption{Annual average of emission per $\text{km}^2$ of sectors for spatiotemporal clusters obtained combining the dissimilarity matrices of the time series of emissions of different sectors between 1990 and 2022, and the geodetic distances across centroids of nuts2 level regions.}
    \label{fig:sp_clu_centroids_sect}
\end{figure}

In Figure \ref{fig:map_sp_clu_sect} we present a map of the regions belonging to different clusters and in Figure \ref{fig:sp_clu_centroids_sect} we report the average annual emission in each cluster, for each sector. 


Cluster 1 primarily includes regions located in Central Europe, such as Germany, Denmark, the Czech Republic, Austria, Italy, Croatia, Slovenia, along with a few regions of France. This cluster exhibits steady levels of emissions, which are neither the highest nor the lowest compared to other clusters across all sectors.

Cluster 2 consists primarily of regions in Eastern Europe, including Finland, Sweden, Latvia, Lithuania, Poland, Slovakia, Hungary, Romania, Bulgaria, and Greece. This cluster exhibits the overall lowest emission levels, with stable trends in the Buildings, Energy, Industry, and Waste sectors. Emissions in the Agricultural sector show a slight decrease during the 1990s, while emissions in the Transport sector display a gradual increase over time.
Cluster 3 includes the smallest number of regions, located in Belgium and the Netherlands and some metropolitan areas in Germany and Eastern Europe. As highly industrialized and densely populated areas, it is unsurprising that this group exhibits the highest levels of emissions across all sectors except Agriculture. In particular, emissions from the Buildings and Industry sectors display a decreasing trend, declining from 6 tonnes per $\text{km}^2$ to less than 4 tonnes per $\text{km}^2$. Emissions from the Energy sector demonstrate high variability over time, with a clear reduction from 6 to 3.5 tonnes per $\text{km}^2$ in the last five years. Waste emissions also follow a generally decreasing trend but appear to have stabilized in recent years. Notably, only the Transport sector shows an increase in emissions over time. 

Considering Cluster 4, it includes regions located in Belgium and the Netherlands, Northen France and Northen Germany. This group is characterized by particularly high levels of emissions in the Agricultural sector, which exhibit a decreasing trend over time. For other sectors, it ranks as the second worst cluster, following Cluster 3. In the Transport and Waste sectors, emissions follow a dynamic similar to Cluster 3, showing increasing and decreasing trends, respectively, while no clear trend is observed in the Buildings, Energy, and Industry sectors.

Finally, cluster 5 comprises regions located in Spain, Portugal, Ireland, and certain areas of France. This group is characterized by relatively low emissions across all sectors. Notably, emission levels in this group do not exhibit pronounced rising or falling trends. However, a slight upward trend is observed in the Transport sector, while the Energy sector shows a modest decline over time. 

Overall, we can conclude that a relatively small number of regions, encompassing a limited portion of the European territory, are responsible for exceptionally high GHG emissions. Within each breakdown, a group with particularly elevated emission levels could be distinctly identified. While it is clear that this small group contributes disproportionately to total emissions, it is equally evident that these regions also exhibit decreasing trends over time. This suggests ongoing improvements, albeit at varying rates, across individual sectors and for gases with higher emission levels.

To further investigate the role of the spatial component in cluster formation, we performed non-spatial clustering as a robustness check for emissions by sector, similarly to the previous case. The resulting clusters show greater overlap compared to the spatiotemporal clustering approach. Nevertheless, regions in Central-Northern Europe and metropolitan areas remain identifiable as those with the highest levels of emissions per $\text{km}^2$ across multiple sectors. A detailed discussion of this robustness check and the associated results can be found in Appendix \ref{App:ts_clu_sector}.


\section{Conclusion}\label{Sec_Conclusions}

Spatiotemporal hierarchical clustering methodology is a versatile and flexible approach to analyze patterns and dynamics of complex phenomena. In our study, we applied this methodology to greenhouse gas emissions across European regions, combining temporal emission patterns and spatial proximity into a unified analysis. Notably, we sought to assess whether the geographical information is helpful in explaining emission dynamics, providing both methodological advancements and empirical insights.

To this end, we introduced a novel measure, the Joint Inertia, to evaluate the role of the spatial component in the clustering process. This measure captures the contribution of spatial distances in generating the final clustering partition. Through a simulation experiment, we demonstrated the advantages of our methodology and highlighted how the Joint Inertia effectively quantifies the influence of spatial information on the clustering results. These findings underscore the robustness of our approach and its capacity to reveal spatial-temporal structures in the data that might otherwise remain hidden.

Our application provides compelling evidence that spatial context meaningfully enhances the analysis of emission trends. The Joint Inertia values indicate that including spatial distances enriches the clustering process, revealing region-specific patterns and dynamics that reflect Europe’s inherent heterogeneity in economic, industrial, and social characteristics. 

By separately analyzing emissions by gas type and sector, we uncovered distinct trends and dynamics within each cluster. This analysis highlighted how regions with high emissions are often associated with specific sectors, such as agriculture or industry, while others demonstrate relatively low emissions with stable or slowly evolving patterns. The interplay of gas-specific and sector-specific emissions varies significantly across regions, reflecting the diverse economic, industrial, and social landscapes of Europe. This detailed approach provides a more nuanced understanding of emission patterns, identifying both hotspots of high emissions and regions with notable improvements or stability. Such insights are invaluable for tailoring regional environmental policies and strategies to address sector-specific challenges effectively.

The methodological contributions and empirical findings of this study hold significant utility for both statisticians and environmental researchers. From a methodological perspective, our approach demonstrates the versatility of spatiotemporal clustering for high-dimensional data, offering a framework that can be adapted to other environmental and socioeconomic phenomena. For environmental research, this study delivers actionable insights into the spatiotemporal dynamics of greenhouse gas emissions, offering a foundation for targeted mitigation policies and resource allocation.

Future work could extend these contributions by linking cluster dynamics to socioeconomic, industrial, or policy variables, further enriching the understanding of emission drivers, exploiting proper spatiotemporal models. This would enable a more comprehensive framework for analyzing greenhouse gas emissions in a holistic spatiotemporal context, paving the way for interdisciplinary research and evidence-based policy development.



\section*{Declarations}
\begin{itemize}
\item Funding: This study was partly funded by the University of Milano-Bicocca in the framework of the Sequestering CARbon through Forests, AgriCulture, and land usE (SCARFACE) research project (2024-ATEQC-0048).
\item Conflict of interest and competing interests: The authors have no competing interests to declare that are relevant to the content of this article.
\item Data and code availability: Data used in the paper are public (source JRC-ARDECO). Since we use only public data, no Special Permission is need to use copyrighted material from other sources (including the Internet). All results presented in this paper can be reproduced using the \texttt{R} statistical software. The codes were developed entirely by the authors. For reproducibility purposes, we attach to the present submission code and data. After the peer-review process, in the case of a positive outcome, all the scripts and the data are made public through a dedicated GitHub webpage.
\item Declaration of Generative AI and AI-Assisted Technologies in the Writing Process: During the preparation of this work, the authors used ChatGPT, a generative AI tool developed by OpenAI, to assist with drafting, rephrasing, and refining the text. After utilizing this tool, the authors carefully reviewed, edited, and revised the content as necessary to ensure accuracy, clarity, and alignment with the objectives of the publication. The authors take full responsibility for the content of this manuscript, including any ideas, analyses, and conclusions presented.

\end{itemize}

\begin{appendices}
\section{Technical details of Joint Inertia}
\label{App:joint_inertia}
In this appendix we focus on the properties of Joint Inertia, in particular we explain why its value varies between $0$ and $1$ and what are the conditions that guarantee it, starting from the simplest case where we have only two matrices of dissimilarity and then extending the reasoning to the case where we want to combine more than two matrices.

\subsection{Joint Inertia for the case $P=2$ dissimilarity matrices}

Let $D_0 = [d_{0,ij} ]_{i,j=1,\dots,n}$ and $D_1 = [d_{1,ij} ]_{i,j=1,\dots,n}$ refer respectively to any distances matrix of variables and the spatial distances matrix considering a sample of $n$ units.
Let us consider a partition $\mathcal{P}_K^{\alpha}$ in $K$ clusters obtained mixing the dissimilarity matrices $D_0$ and $D_1$ with the parameter $\alpha$. Also, let us denote its within-clusters mixed inertia as $W(\mathcal{P}_K^{\alpha})$ and the corresponding proportion of the total pseudo inertia explained as
\begin{equation*}
Q_{0}(\mathcal{P}_K^{\alpha})=1-\frac{W_{0}(\mathcal{P}_K^{\alpha})}{W_{0}(\mathcal{P}_1)} \qquad Q_{1}(\mathcal{P}_K^{\alpha})=1-\frac{W_{1}(\mathcal{P}_K^{\alpha})}{W_{1}(\mathcal{P}_1)},
\end{equation*}
where $W_{0}(\mathcal{P}_1)$ and $W_{1}(\mathcal{P}_1)$ are the total pseudo inertia under dissimilarity matrix $D_0$ and under dissimilarity matrix $D_1$, respectively. 
It is easy to observe that the possible values of this ratio vary between $0$ and $1$ because the sum of the square dissimilarities between the units belonging to the same clusters will not be greater than the total of the square dissimilarities within a matrix, and being quantities square, will all have positive values, for each possible value of $\alpha$.

The resulting mixed dissimilarity matrix $D=(1-\alpha) D_0+\alpha D_1$ will reflect more the dissimilarities in $D_0$ (or $D_1$), when the value of $\alpha$ is closer to $0$ (or close to $1$), thus the resulting partition $\mathcal{P}_K^{\alpha}$ will be able to capture an higher proportion of explained inertia in $D_0$ (or $D_1$). Therefore, when $\alpha$ increases, $Q_{D_0}(\mathcal{P}_K^{\alpha})$ will increase and $Q_{D_1}(\mathcal{P}_K^{\alpha})$ will decrease. This is also explained in details in Section 3.2 of \cite{Chavent2018}. Notice that the proportion of explained inertia is equal to $0$ only when $K=1$ and it is $1$ only when $K=n$, for each value of $\alpha$. Following we summarize and formalize the concept which hold for $1<K<n$
\begin{equation*}
0=Q_{D_0}(\mathcal{P}_{K=1}) < Q_{D_0}(\mathcal{P}_K^{\alpha=1}) \leq Q_{D_0}(\mathcal{P}_K^{\alpha}) \leq Q_{D_0}(\mathcal{P}_K^{\alpha=0}) < Q_{D_0}(\mathcal{P}_{K=n}) = 1
\end{equation*}
\begin{equation*}
0=Q_{D_1}(\mathcal{P}_{K=1}) < Q_{D_1}(\mathcal{P}_K^{\alpha=0}) \leq Q_{D_1}(\mathcal{P}_K^{\alpha}) \leq Q_{D_1}(\mathcal{P}_K^{\alpha=1}) < Q_{D_1}(\mathcal{P}_{K=n}) = 1.
\end{equation*}

Now, we consider the normalized proportion of explained inertia for dissimilarity matrix $D_0$ (or $D_1$), which correspond to the ratio between the proportion of explained inertia in the partition obtained using the combined dissimilarity matrix with parameter $\alpha$, and the proportion of inertia explained using only the features (or spatial) dissimilarities, thus the maximum proportion of explained inertia.
\begin{equation*}
\tilde{Q}_{D_0}(\mathcal{P}_K^{\alpha})=\frac{Q_{D_0}(\mathcal{P}_K^{\alpha})}{Q_{D_0}(\mathcal{P}_K^0)} \qquad \tilde{Q}_{D_1}(\mathcal{P}_K^{\alpha})=\frac{Q_{D_1}(\mathcal{P}_K^{\alpha})}{Q_{D_1}(\mathcal{P}_K^1)}.
\end{equation*}

These quantities are strictly positive and no greater than $1$. In particular, the normalized proportion of explained inertia in $D_0$ (or $D_1$) reaches its minimum value when the feature (or spatial) dissimilarity matrix is not included in the combination, thus $\alpha=1$ (or $\alpha=0$), and it reaches its maximum when the clustering partition is obtained from the purely no-spatial (or spatial) case.
\begin{equation*}
0 < \tilde{Q}_{D_0}(\mathcal{P}_K^{\alpha=1}) \leq \tilde{Q}_{D_0}(\mathcal{P}_K^{\alpha}) \leq \tilde{Q}_{D_0}(\mathcal{P}_K^{\alpha=0})=1
\end{equation*}
\begin{equation*}
0 < \tilde{Q}_{D_1}(\mathcal{P}_K^{\alpha=0}) \leq \tilde{Q}_{D_1}(\mathcal{P}_K^{\alpha}) \leq \tilde{Q}_{D_1}(\mathcal{P}_K^{\alpha=1})=1
\end{equation*}

The Joint Inertia between two dissimilarity matrices is computed using the following formula
\begin{equation*}
JI(\mathcal{P}_K^{\alpha})=\tilde{Q}_{D_1}(\mathcal{P}_K^{\alpha})-(1-\tilde{Q}_{D_0}(\mathcal{P}_K^{\alpha}))=\frac{{Q}_{D_1}(\mathcal{P}_K^{\alpha})}{{Q}_{D_1}(\mathcal{P}_K^{1})}+\frac{{Q}_{D_0}(\mathcal{P}_K^{\alpha})}{{Q}_{D_0}(\mathcal{P}_K^{0})}-1.
\end{equation*}
It is strictly positive because the partition obtained from the mixed dissimilarity matrix allows at least to explain as much as the partition obtained from purely non-spatial case in $D_0$ or the partition obtained from purely spatial case in $D_1$. This has already been explained extensively in Section \ref{sec_jointInertia}, through Figure \ref{fig:clusters_cases} and Table \ref{tab:cluster_cases}.

\subsection{Joint Inertia for the case $P>2$ dissimilarity matrices}

The generalization to the case of multiple dissimilarity matrices, one of which relates to the spatial component, is straightforward. Let us suppose that we are interested in computing the Joint Inertia for the generic variable $m$ (with $m$ being the generic index for the $P$ variables of interests) compared to the combination of all the other variables used to cluster the time series \footnote{In the main manuscript, we focused on the specific case of the Joint Inertia for the spatial dissimilarity matrix, that we assumed to be the $P$-th variable. Hence, the results in the Appendix can be related to the results in the main manuscript by assuming that $m=P$.}.

Recalling the notation previously introduced, let us consider a set of $P$ variables observed over time at $n$ spatial locations (or regions) and let $\underline{\alpha}=[\alpha_p]_{p=1,2,\dots,P}=(\alpha_1,\alpha_2,\ldots,\alpha_p,\ldots,\alpha_P)$ be the vector of mixing weights\footnote{Notice that, although the grid can be either regular or irregular, here we consider a regular grid with constant step $\Delta \alpha$.} used to compute the linear combination of all the dissimilarity matrices $D(\underline{\alpha})=\sum_{p=1}^P \alpha_p D_p$ subject to $\alpha_p>0 \quad \forall p=1,\ldots,P$ and $\sum_{p=1}^P \alpha_p = 1$.

By applying the Ward-like hierarchical clustering algorithm presented in the manuscript to the matrix $D(\underline{\alpha})$ generated by a given combination of mixing parameters $\underline{\alpha}$ and number of clusters $K$ the resulting partition is denoted as $\mathcal{P}_K^{\underline{\alpha}}$. Given the generic partition $\mathcal{P}_K^{\underline{\alpha}}$ obtained from the mixed dissimilarity matrix, the proportion of explained inertia for matrix $m$-th is computed as
\begin{equation*}
Q_{D_p}(\mathcal{P}_K^{\underline{\alpha}})=1-\frac{W_{D_p}(\mathcal{P}_K^{\underline{\alpha}})}{W_{D_p}(\mathcal{P}_1)} 
\end{equation*}
and the normalized proportion of explained inertia, with respect to the case in which only variable $p$ is used, is computed as
\begin{equation*}
\tilde{Q}_{D_p}(\mathcal{P}_K^{\underline{\alpha}})=\frac{Q_{D_p}(\mathcal{P}_K^{\underline{\alpha}})}{Q_{D_p}(\mathcal{P}_K^{\alpha_p=1})}
\end{equation*}
Recall that these quantities are both included between $0$ and $1$ for the same reasons and considerations as in the case of two dissimilarity matrices.

To retrieve the equation of the Joint Inertia for variable $m$ we need to focus on two dissimilarity matrices: the first is the matrix related to $m$-th variable of interest on which to compute the index and the second is the linear combination of all the others dissimilarity matrices different from $m$ (i.e., the complementary matrices). Let $D_m = [d_{m,ij}]_{i,j=1,\dots,n}$ be the dissimilarity matrix for variable $m$ across the $n$ units and let $D_{-m} = [d_{p,ij}]_{p=1,\dots,m-1,m+1,\ldots,P; i,j=1,\dots,n}$ be the dissimilarity matrices containing the distances across the $P-1$ time series different from $m$. Moreover, let use define the matrix $D_{-m}(\underline{\alpha}|\alpha_m=0)=\sum_{p \neq m} \alpha_p D_p$ as the combination of the $P-1$ dissimilarity matrices excluding the $m$-th dissimilarity matrix $D_m$, and the resulting clustering partition as $\mathcal{P}_K^{\underline{\alpha}|\alpha_m=0}$. Similarly, define the partition obtained by using only the information contained in the $m$-th matrix (i.e., by fixing $\alpha_m=1$) as $\mathcal{P}_K^{\alpha_m=1}$.

If we denote the total inertia in $D_{-m}$ as $W_{D_{-m}}(\mathcal{P}_1)$, we can compute the following quantities
\begin{equation*}
Q_{D_{-m}}(\mathcal{P}_K^{\underline{\alpha}})=1-\frac{W_{D_{-m}}(\mathcal{P}_K^{\underline{\alpha}})}{W_{D_{-m}}(\mathcal{P}_1)} \qquad Q_{D_{-m}}(\mathcal{P}_K^{\underline{\alpha}|\alpha_m=0})=1-\frac{W_{D_{-m}}(\mathcal{P}_K^{\underline{\alpha}|\alpha_m=0})}{W_{D_{-m}}(\mathcal{P}_1)}
\end{equation*}
which represent the proportion of explained inertia induced by the complementary matrices evaluated at the generic vector of mixing parameters $\underline{\alpha}$ (left formula) and the the proportion of explained inertia induced by the complementary matrices evaluated at the restricted vector $\underline{\alpha}|\alpha_m=0$ which does not consider the $m$-th matrix\footnote{Notice that, when $m$ is the spatial matrix, the first formula corresponds to the proportion of explained inertia induced by the feature combined dissimilarity matrix when including the spatial component, while the second formula is the proportion of explained inertia induced by the feature combined dissimilarity matrix when excluding the spatial component.} (right formula).

The normalized proportion of explained inertia in the non-spatial combined dissimilarity matrix can be computed as 
\begin{equation*}
\tilde{Q}_{D_{-m}}(\mathcal{P}_K^{\underline{\alpha}})=\frac{Q_{D_{-m}}(\mathcal{P}_K^{\underline{\alpha}})}{Q_{D_{-m}}(\mathcal{P}_K^{\underline{\alpha}^*|\alpha_m=0})}   
\end{equation*}

By definition, all these quantities are bounded between $0$ and $1$. Indeed, If we employ one of the procedures in \cite{Chavent2018} or \cite{Morelli2024} to compute the optimal weighting vector for the combination of all the dissimilarity matrices different from $m$ (i.e., $\underline{\alpha}^*|\alpha_m=0$ subject to $\sum_{p \neq m} \alpha_p = 1$), then, the corresponding convex combination of dissimilarity matrices (i.e., $D_{-m}(\underline{\alpha}^*|\alpha_m=0)$) will explain as much as possible the dissimilarities in the $D_m$ matrices, thus being $Q_{D_{-m}}(\mathcal{P}_K^{\underline{\alpha}^*|\alpha_m=0})$ greater than any other potential $Q_{D_{-m}}(\mathcal{P}_K^{\underline{\alpha}|\alpha_m=0})$.

Finally, considering a generic combination of $\underline{\alpha}$ and $K$, the Joint Inertia for variable $m$ when the number of dissimilarity matrices is $P>2$ can be determine by the following formula
\begin{equation*}
JI_m(\mathcal{P}_K^{\underline{\alpha}})=\tilde{Q}_{D_m}(\mathcal{P}_K^{\underline{\alpha}})-(1-\tilde{Q}_{D_{-m}}(\mathcal{P}_K^{\underline{\alpha}}))=\frac{Q_{D_m}(\mathcal{P}_K^{\underline{\alpha}})}{Q_{D_m}(\mathcal{P}_K^{\alpha_m=1})}+\frac{Q_{D_{-m}}(\mathcal{P}_K^{\underline{\alpha}})}{Q_{D_{-m}}(\mathcal{P}_K^{\underline{\alpha}^*|\alpha_m=0})}-1.
\end{equation*}

Similarly to the case where we had only two matrices, if the matrix $D_m$ does not play some useful role in identifying groups compared with all other dissimilarity matrices, one of the normalized proportion of explained inertia takes value $1$ and the other one will be close to $0$ but strictly positive. If the dissimilarity matrix $D_m$ alone and the combined dissimilarity matrix excluding $D_m$ lead exactly to the same partition we would obtain using the combination of all dissimilarity matrices, thus $\mathcal{P}_K^{\alpha_m=1} = \mathcal{P}_K^{\underline{\alpha}^*|\alpha_m=0} = 
\mathcal{P}_K^{\underline{\alpha}} $ the resulting Joint Inertia is exactly $1$, thus its maximum value, reflecting the crucial role of matrix $D_m$ which is able to perfectly explain the same sum of square dissimilarity of the other variables, even if they are not included.

\section{Non-spatial clustering}
\label{App:ts_clu}

With the aim of better illustrating the utility of the spatiotemporal clustering methodology employed and emphasizing the role of the spatial component—central to this paper—in this appendix, we present the results obtained from performing the cluster analysis without incorporating the spatial component. This comparison underscores the added value of integrating spatial information in identifying meaningful patterns and dynamics in the data.

\subsection{Temporal clustering for emissions by gas}\label{App:ts_clu_gas}

\begin{figure}
    \centering
    \includegraphics[width=1\linewidth]{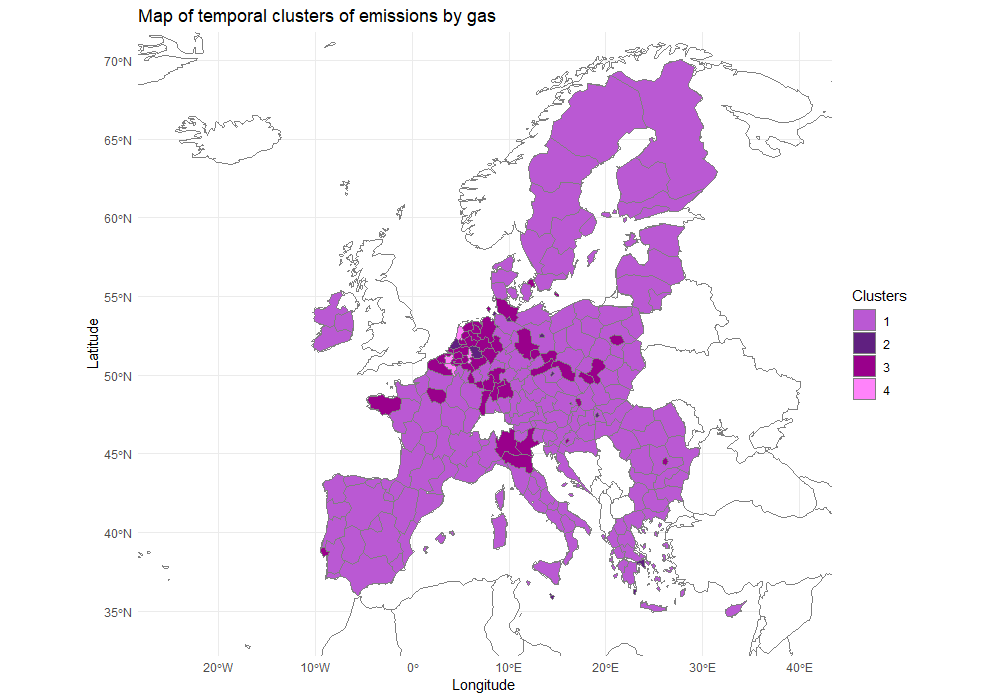}
    \caption{Map of temporal clusters obtained combining the dissimilarity matrices of the time series of emissions of different gases between 1990 and 2022.}
    \label{fig:map_ts_clu_gas}
\end{figure}

\begin{figure}
    \centering
    \includegraphics[width=1\linewidth]{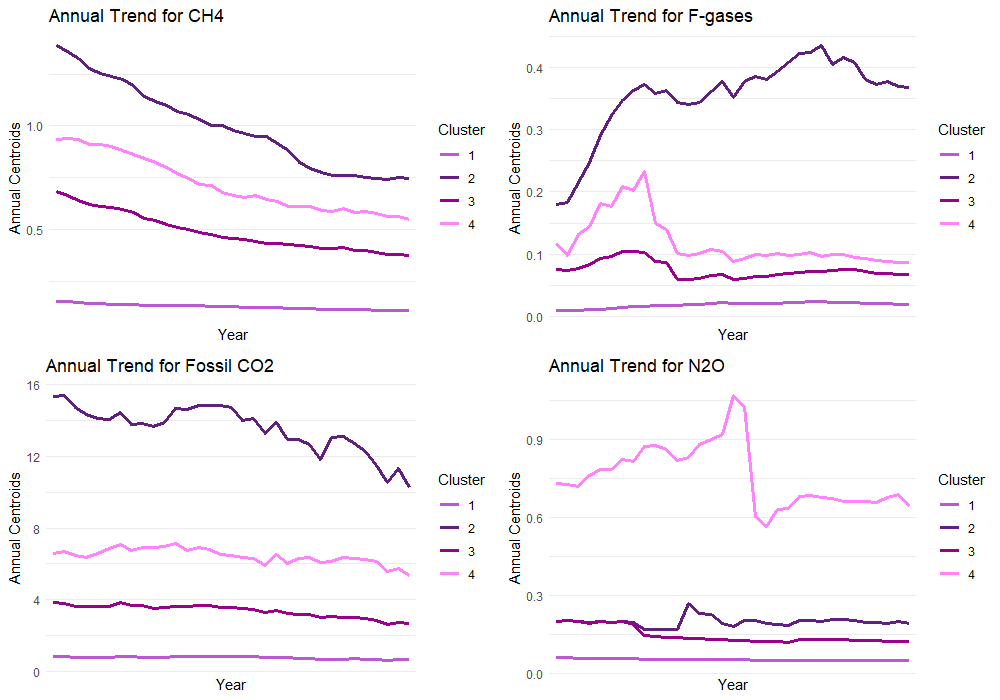}
    \caption{Annual average of emission per $\text{km}^2$ of sectors for temporal clusters obtained combining the dissimilarity matrices of the time series of emissions of different gases between 1990 and 2022.}
    \label{fig:ts_clu_centroids_gas}
\end{figure}

Considering dissimilarity matrices $D_{CH_4}$, $D_{\text{F-gases}}$, $D_{N_2O}$ and $D_{CO_2}$, as defined in Section \ref{sec:spt_gas}, we set $\Delta \alpha=0.05$ and $K=4$ and we obtain ${\alpha_{4,p}}^*=(0.1,0.35,0.15,0.4)$ as the optimal weighting vector for the dissimilarity matrices.
In Figure \ref{fig:map_ts_clu_gas} we show a map of the regions belonging to different clusters and in Figure \ref{fig:ts_clu_centroids_gas} we provide the average annual emission in each cluster by gas.

The clustering results exhibit a high degree of spatial overlap, yet discernible spatial patterns remain evident. Notably, regions in clusters 1 and 2, which demonstrate the poorest performance, include Belgium, the Netherlands, and certain metropolitan areas in Northern Europe. These regions display a decreasing trend in emissions for the gases with the highest levels, namely $CH_4$ and $CO_2$. In contrast, cluster 3 consists of sparsely populated regions in Central Europe with significantly lower emissions than the previous clusters. Lastly, the cluster encompassing the largest number of regions, covering over two-thirds of the analyzed area, is characterized by remarkably low emissions and appears relatively stable over time.

\subsection{Temporal clustering for emissions by sector}\label{App:ts_clu_sector}

\begin{figure}
    \centering
    \includegraphics[width=1\linewidth]{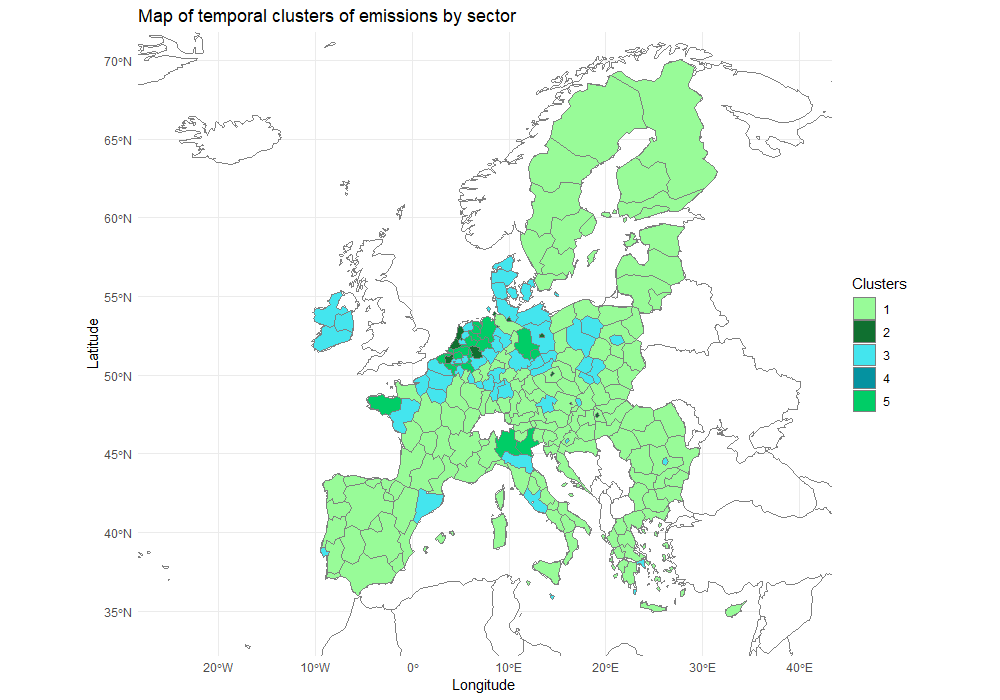}
    \caption{Map of temporal clusters obtained combining the dissimilarity matrices of the time series of emissions of different sectors between 1990 and 2022.}
    \label{fig:map_ts_clu_sect}
\end{figure}

\begin{figure}
    \centering
    \includegraphics[width=1\linewidth]{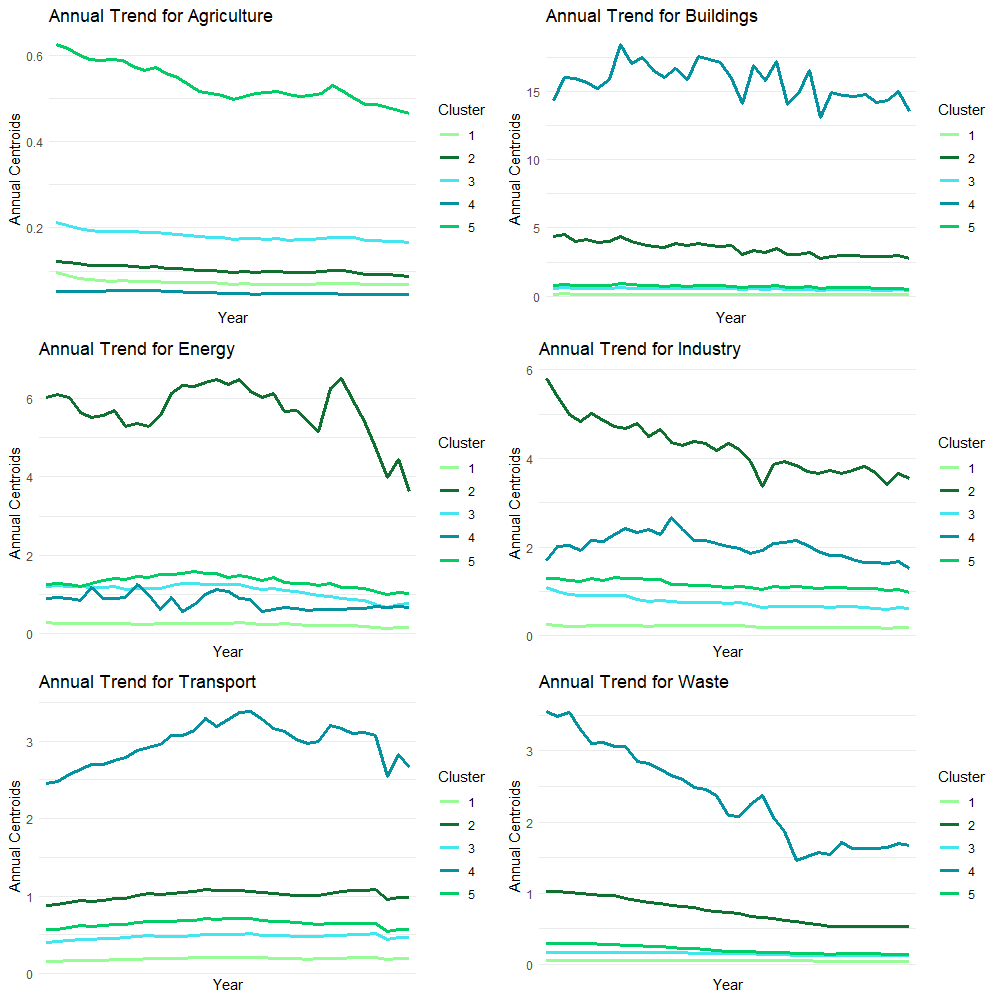}
    \caption{Annual average of emission per $\text{km}^2$ of sectors for temporal clusters obtained combining the dissimilarity matrices of the time series of emissions of different sectors between 1990 and 2022.}
    \label{fig:ts_clu_centroids_sect}
\end{figure}

Moving to the emissions by sectors, we consider dissimilarity matrices $D_{\text{Agri}}$, $D_{\text{Build}}$, $D_{\text{Energy}}$, $D_{\text{Ind}}$, $D_{\text{Tran}}$ and $D_{\text{Waste}}$, as defined in Section \ref{sec:spt_sector}, we set $\Delta \alpha=0.1$ and $K=5$ and we obtain ${\alpha_{5,p}}^*=(0.3,0,0.2,0,0.3,0.2)$ as the optimal weighting vector for the dissimilarity matrices.
In Figure \ref{fig:map_ts_clu_sect} we provide a map of the regions belonging to different clusters and in Figure \ref{fig:ts_clu_centroids_sect} we describe the average annual emission in each cluster by gas.

From the results, we can immediately see that, even without incorporating the spatial component, there is still a degree of spatial overlap in the clustering outcomes. Notably, regions with particularly low performance are concentrated within the European Union. Specifically, Cluster 1, the largest in terms of geographic area, exhibits very low emission levels across all sectors, with no discernible increasing or decreasing trends. Cluster 2 encompasses industrial regions in Northern Europe, characterized by high emissions in the Energy and Industry sectors, which show a declining trend over time. Cluster 3 is associated with low, though not minimal, emission levels, accompanied by a slight downward trend. Cluster 4 includes several metropolitan regions and displays the highest emission levels in the Buildings, Transport, and Waste sectors. Lastly, Cluster 5, primarily located in Northern Europe and Northern Italy, stands out as the largest emitter in the Agricultural sector.

Overall, our analysis confirms the significant relevance of the spatial component. Regions that are geographically close tend to exhibit similar emission patterns, both by gas type and sector. Incorporating the spatial component explicitly into the clustering algorithm produces groups that are more interpretable in terms of geographic proximity. This leads to results that are particularly valuable in environmental statistics, where spatiotemporal approaches and methodologies are essential for understanding and addressing complex phenomena.

\end{appendices}

\bibliography{sn-article}

\end{document}